\documentclass[aps,amsmath,amssymb,preprintnumbers,groupedaddress,twocolumn,nofootinbib]{revtex4-1}
\usepackage{tikz} \usepackage{longtable}
\usepackage{lipsum}

\newcommand{\drawsquare}[2]{\hbox{%
\rule{#2pt}{#1pt}\hskip-#2pt
\rule{#1pt}{#2pt}\hskip-#1pt
\rule[#1pt]{#1pt}{#2pt}}\rule[#1pt]{#2pt}{#2pt}\hskip-#2pt
\rule{#2pt}{#1pt}}
\newcommand{\fund}{\raisebox{-.5pt}{\drawsquare{6.5}{0.4}}}
\newcommand{\ov}{\overline}

\newcommand{\bF}{\mathbb{F}}
\newcommand{\F}{\mathbb{F}}
\newcommand{\bZ}{\mathbb{Z}}
\newcommand{\Z}{\mathbb{Z}}
\newcommand{\Q}{\mathbb{Q}}
\newcommand{\C}{\mathbb{C}}
\newcommand{\bC}{\mathbb{C}}
\newcommand{\bP}{\mathbb{P}}

\newcommand{\yy}{u}
\newcommand{\YY}{U}
\definecolor{Green}{RGB}{0,200,0}

\def\P{\mathbb{P}}
\def\E{\mathbb{E}}

\newcommand{\sskip}{\vspace{.5cm}}
\newcommand{\vsskip}{\vspace{.1cm}}

\newcommand{\cN}{\mathcal{N}}

\newcommand{\cO}{\mathcal{O}}

\newcommand{\lqcd}{\Lambda_\text{QCD}}
\newcommand{\shs}{\hspace{.2cm}}

\begin{document}

\preprint{NSF-KITP-14-100, MIT-CTP-4581}

\title{Non-Higgsable QCD and the Standard Model Spectrum in F-theory}

\author{Antonella Grassi$^1$, James Halverson$^2$, Julius Shaneson$^1$, and
  Washington Taylor$^3$ \vspace{.3cm}} 

\affiliation{$^1$Department of Mathematics, University of Pennsylvania,
  Philadelphia, PA 19104} 
\affiliation{$^2$Kavli Institute for Theoretical
  Physics, University of California, Santa Barbara, CA 93106 USA}
\affiliation{$^3$Center for Theoretical Physics, Massachusetts Institute
  of Technology, Cambridge, MA 02139, USA}

\begin{abstract}
\noindent 
Many four-dimensional supersymmetric compactifications of F-theory
contain gauge groups that cannot be spontaneously broken through
geometric deformations.  These ``non-Higgsable clusters'' include
realizations of $SU(3)$, $SU(2)$, and $SU(3) \times SU(2)$, but no
$SU(n)$ gauge groups or factors with $n> 3$.  We study possible
realizations of the standard model in F-theory that utilize
non-Higgsable clusters containing $SU(3)$ factors and show that there
are three distinct possibilities.  In one, fields with the non-abelian
gauge charges of the standard model matter fields  are localized at
a single locus where non-perturbative $SU(3)$ and $SU(2)$
seven-branes intersect; cancellation of gauge anomalies implies that
the simplest four-dimensional chiral $SU(3)\times SU(2)\times U(1)$
model that may arise in this context exhibits standard model families.  We
identify specific geometries that realize non-Higgsable $SU(3)$ and
$SU(3) \times SU(2)$ sectors.  This kind of scenario provides a
natural mechanism that could explain the existence of an unbroken QCD
sector, or more generally the appearance of light particles and
symmetries at low energy scales.
\end{abstract}

\maketitle

\section{Introduction}

In recent years there has been much progress in understanding the
physics of F-theory \cite{Vafa-F-theory, Morrison-Vafa-I,
  Morrison-Vafa-II} compactifications.  An F-theory compactification
to $d=4,6$ or $8$ space-time dimensions can be thought of as a
supersymmetric type IIB compactification with a
varying axiodilation profile $\tau = C_0 + i/g_s$; another definition
that is more precise for some purposes arises by considering an
M-theory compactification to $d=3,5,$ or $7$ dimensions on an
elliptically fibered Calabi-Yau manifold in a vanishing fiber limit
that corresponds to a decompactification limit via a single T-duality.
The set of F-theory compactifications represents a promising region of
the landscape for a number of reasons. One is that grand unified
models in F-theory have some features that are more realistic than
their weakly coupled type II counterparts --- though as we show in
this paper, even in models without grand unification F-theory has
certain desirable phenomenological features not present in the weakly
coupled limit. Perhaps most importantly, F-theory appears to provide
the broadest view currently available of the landscape of ${\cal N}=
1$ string compactifications to six or four dimensions.  See
\cite{Morrison-TASI, Denef-F-theory, WT-TASI} and
\cite{Heckman-review, Weigand-review} respectively for pedagogical and
phenomenological
reviews of F-theory.

An interesting feature of many string compactifications is the
presence of ``non-Higgsable clusters'' \cite{clusters}.
These are connected
gauge sectors, carried by seven-branes in F-theory, that cannot be
spontaneously broken.  The simplest non-Higgsable clusters consist of
a single gauge group with little or no charged matter.  It has been
known since the early days of F-theory that certain gauge groups,
particularly $SU(3), SO(8), F_4, E_6, E_7,$ and $E_8$, can be realized
through dual heterotic and F-theory compactifications in models with no
charged matter, so that these groups cannot be Higgsed
\cite{Morrison-Vafa-II}.  Non-Higgsable clusters can also contain
multiple gauge group factors and charged matter.  A systematic
classification of these clusters for six-dimensional F-theory models
was given in \cite{clusters}, and includes non-Higgsable gauge group
products such as $G_2 \times SU(2)$ with jointly charged matter.

Such non-Higgsable clusters with multiple gauge group factors and
matter are unique to F-theory in known string constructions.
Mathematically, non-Higgsable clusters are produced when the
elliptically fibered Calabi-Yau manifold $X$ used for an F-theory
compactification has certain types of singularities at generic points
in its complex structure moduli space $CS(X)$.  Physically, from the
point of view of the corresponding low-energy theory, a non-Higgsable
structure arises when there is no symmetry-breaking flat direction in
the supersymmetric moduli space.  For six-dimensional theories, there
is a direct correspondence between the geometric structure and the
non-Higgsability of the low-energy theory.

We emphasize at the outset that the situation is more complicated
in four-dimensional theories, since there are additional features that
affect the low-energy physics. In particular, though a
four-dimensional F-theory compactification with a geometrically
non-Higgsable cluster exhibits seven-brane gauge sectors that cannot
be broken by complex structure deformation, we cannot rule out the
possibility of other effects that may, in certain cases, break the
gauge group.  These include additional moduli and a set of discrete
fluxes; the latter corresponds to $G_4$ flux in the M-theory picture,
which we henceforth refer to as G-flux in the F-theory picture. There
are often many discrete flux choices (in fact for some $X$ the G-flux
must always be non-trivial), and depending on the choice there are at
least three possible effects on the low-energy theory: it may induce a
flux-breaking of the gauge group, give rise to a chiral spectrum,
and/or stabilize some of the complex structure moduli of $X$. Though
we focus primarily on the geometry of $X$ and its complex structure,
the scenario we propose operates under the assumptions that one does
not turn on fluxes that break the geometrically imposed gauge group
$G$ (as typically assumed in the F-theory GUT literature) or force an
enhancement of $G$ through moduli stabilization on a locus with
non-generic gauge group.  We discuss these issues in section
\ref{sec:issues-G-flux}, where we also present a discussion of
the relationship between geometric and physical non-Higgsability in
four-dimensional theories.  Throughout this paper we identify and
study geometrically non-Higgsable clusters, which correspond to
seven-brane sectors that cannot be broken by complex structure
deformation; in principle we expect that these will generally
correspond to non-Higgsable structure in the resulting low-energy 4D
supergravity theory, but in some cases this relationship may be
subject to the caveats above.

The complete set of gauge group factors that can be associated to a
non-Higgsable geometry is rather restricted.  In decreasing order of
dimension, the possible simple (or abelian)
factors $G_i$ in a
non-Higgsable gauge group $G=\prod_i G_i$ (in {\it any} dimension)
are in the set:\footnote{Note that  the actual gauge group can
  also involve a quotient by a discrete group, so that strictly
  speaking one should say that the set above gives the complete list of
  possible gauge algebras.  In this paper we do not worry about the
  global structure of the group, and simply refer to the various
  simple group factors.}
\begin{align}
 \{
E_8, E_7, E_6, F_4, SO(8), SO(7),  
G_2,
  SU(3), SU(2), U(1)\}.
\end{align}
Notably, $SU(5)$ and $SO(10)$ are both absent from this list and
$SU(3)$ and $SU(2)$ are the only possible $SU(n)$ groups.  The
singularities that may give rise to a non-Higgsable $SU(3)$ or $SU(2)$
(or any other non-Higgsable gauge group factor)
do not admit a description in terms of perturbative string theory on
$D7$ branes; they arise from non-trivial $(p,q)$ seven-branes.

Though in six dimensions the only non-Higgsable
product groups that can arise are
$G_2 \times SU(2)$ and $SU(2) \times SO(7) \times SU(2)$ (again, up to
possible quotients by a discrete subgroup), in four dimensions the set
of possibilities is richer. As we show in this paper, this includes
the possibility of an $SU(3) \times SU(2)$ non-Higgsable cluster or a
non-Higgsable $SU(3)$ with Higgsable $SU(2)$.  In both cases, we will
see that the specific spectrum of charged matter exhibited by the
geometry is relevant for particle physics.  Regarding the first of
these possibilities, while of course the $SU(2)$ of the standard model
is in fact broken in the infrared by the Higgs field itself, in some
reasonable scenarios this occurs through radiative electroweak
symmetry breaking and may still be compatible with a non-Higgsable
$SU(2)$ in the model with unbroken SUSY.

It is worth emphasizing that the same local singularity types that can
give non-Higgsable seven-branes carrying specific gauge group factors
can also arise in other geometries from Higgsable seven-brane
configurations carrying the same gauge group factors but with (in general)
more
charged matter; in contrast, there are other singularities, and
associated gauge group factors, that may never be non-Higgsable.  From
the F-theory point of view, many different gauge groups with a rich
variety of possible matter spectra can be arranged by tuning the
axiodilaton profile $\tau$ over a given compactification space; this
corresponds geometrically to tuning the structure of the elliptically
fibered Calabi-Yau manifold by varying the parameters in a Weierstrass
model, as described in more detail in the next section.  Such tuning,
which can give rise to (Higgsable versions of) the gauge group types
mentioned above as well as many other types including general $SU(n)$
and $SO(n)$ factors, has been the primary approach taken
to F-theory model building to date.

\vspace{.5cm}
In this paper we give a systematic
description of the different possible ways that the nonabelian $SU(3)$
and $SU(2)$ factors in the standard model can be realized using
geometrically non-Higgsable and/or Higgsable structures.  We focus in
particular on possible F-theory realizations of the standard model
that utilize a non-Higgsable $SU(3)$ factor; there are three
qualitatively different possibilities in this case for how the $SU(2)$
factor is realized.

Constructions with such a non-Higgsable QCD sector yield two important
physical features.  The first is that the type $IV$ singular fiber
that realizes a non-Higgsable $SU(3)$ factor gives rise to a rich
matter spectrum when it intersects singularity carrying an
$SU(2)$ gauge group; intuitively, this is because the type $IV$ fiber
(without monodromy) realizes an $SU(3)$ gauge theory on \emph{four}
$(p,q)$ seven-branes, as opposed to an $SU(3)$ gauge factor produced
from three D7-branes at weak coupling.  We find in particular that
fields carrying all of the representations of $SU(3) \times SU(2)$
needed for the complete set of standard model matter matter fields may
be localized at the $SU(3)\times SU(2)$ intersection obtained by the
intersection of a type IV and type III fiber; this is in sharp contrast
to the type IIB $SU(3)\times SU(2)$ intersection of two stacks of
D7-branes, which realizes only bifundamental matter that has the
quantum numbers of quark doublets.

The second potentially interesting feature of a non-Higgsable $SU(3)$
factor is that it provides a natural mechanism that could explain the
existence of an unbroken QCD sector in nature.  While this is not an
issue in the standard model itself, the necessary existence of
electromagnetically charged and/or colored scalars in supersymmetric
theories allows for the possible existence of charge or color-breaking
(CCB) vacua in supersymmetric extensions of the standard model.  In
the MSSM the absence of dangerous CCB vacua places bounds on soft
supersymmetry breaking terms, which can rule out or significantly
constrain specific models; this issue is discussed further in Section
\ref{sec:discussion}.

More broadly, the constructions we describe here may have some
relevance for the question of why nature provides us with symmetries
and light particles at low energy scales. In both field theory and
string theory, mechanisms have been proposed for moduli stabilization
at enhanced symmetry points; see e.g. \cite{Beauty} and references
therein.  Such mechanisms are motivated in part by the presence of an
unbroken QCD sector with a confinement scale far below the Planck mass
and the common expectation that typical vacua in the landscape break
gauge symmetry at high scale.  In much of the literature in this area,
in fact, enhanced symmetry points are assumed to be relatively rare
and in particular non-generic in the supersymmetric moduli space.  If
this assumption is true, then indeed some mechanism would be
desirable to explain why vacua are stabilized at those special points
in the moduli space that have enhanced symmetries; the authors of
\cite{Beauty} and other related works also seek to develop mechanisms
that explain the cosmological dynamics that might drive the theory to
these vacua.

Notably, F-theory compactifications with non-Higgsable clusters
sidestep much of this issue, since they exhibit gauge symmetry at
\emph{generic} points in their supersymmetric moduli space.  In
particular, in the kind of QCD scenarios that we describe here, the
existence of an unbroken QCD sector does not require stabilization at
a special locus in moduli space: a generic point will suffice. Though
conventional $SU(5)$ grand unification is impossible in this scenario,
it is particularly intriguing that the theory singles out $SU(3)$ and
$SU(2)$ from all other $SU(n)$ and leads naturally toward the spectrum
of the standard model.

This paper is organized as follows.  In Section
\ref{sec:F-theory-clusters} we review some basics of F-theory and
non-Higgsable clusters.  Section \ref{sec:SM} gives a systematic
description of the possible ways that any F-theory model may be
constructed that contains within it the $SU(3) \times SU(2) \times
U(1)$ of the standard model gauge group.  In Section
\ref{sec:nonhiggsableQCD} we introduce non-Higgsable QCD and its three
qualitatively different realizations. Utilizing a string junction
analysis, we study the seven-branes associated with Kodaira type $IV$
and $III$ singularities, and also the structure of a $IV$-$III$
intersection, which turns out to \emph{generically} involve an extra brane
carrying an $I_1$ singular fiber. Section \ref{sec:examples} contains
a variety of examples of specific F-theory compactifications to 6D and
4D that illustrate the general ideas of the paper and serve as
existence proofs for the various types of non-Higgsable $SU(3)$
models; these examples could be analyzed in more detail in future
work.  In Section \ref{sec:issues} we discuss some of the technical
issues that must be addressed to describe the detailed structure of
low-energy supergravity theories that incorporate a geometrically
non-Higgsable $SU(3)$.

The reader who is interested in our results, but may not be as
familiar with F-theory, may find it useful to read the final section
(\S\ref{sec:discussion}) first.  In this concluding section, we
summarize our results on the three qualitatively different possible
realizations of non-Higgsable QCD and show that the cancellation of
gauge anomalies implies that the simplest four-dimensional chiral
$SU(3)\times SU(2)\times U(1)$ model arising from a non-Higgsable QCD
scenario exhibits standard model families. We also discuss the
physical implications of this scenario, in particular with regard to
model building and enhanced symmetry points in the string landscape.

\section{F-theory, Non-Higgsable Clusters, and Seven-Branes}
\label{sec:F-theory-clusters}

Let us briefly review some relevant facts about F-theory, $(p,q)$
seven-branes, and non-Higgsable clusters.

An F-theory compactification is most easily thought of as a
generalization of type IIB string theory.  Although the M-theory
description is more complete for some purposes, we will focus on the
physics of seven-branes, which are determined geometrically by
promoting the type IIB axiodilaton $\tau = C_0 + i/g_s$ to be the
complex structure modulus of an elliptic curve ({\it i.e.}, a complex
torus with a marked point).  A supersymmetric F-theory
compactification is defined by compactifying type IIB on a complex
manifold $B$ with an axiodilaton profile that defines an elliptically
fibered Calabi-Yau manifold $X$, where a projection $\pi: X
\rightarrow B$ gives the fibration, with $\pi^{-1} (p)=\E$ an elliptic
curve at each point on $B$.  Though the K\" ahler modulus of the
elliptic fiber vanishes in the F-theory limit, its complex structure
remains intact and determines the structure of seven-branes.  The
elliptic fiber can become degenerate along codimension one loci that
mark the presence of type IIB seven-branes, and the axiodilaton
undergoes a nontrivial $SL(2,\Z)$ monodromy along a curve that goes
around such a seven-brane.  For us it will be critical that F-theory
contains a broader set of possible seven-branes than the D7-branes and
O7-planes of the weakly coupled type IIB theory; for example,
seven-brane configurations can arise that carry exceptional gauge
groups and non-perturbative realizations of $SU(3)$ and $SU(2)$.

Seven-branes associated with localized singularities in the function
$\tau$ on $B$ determine many aspects of the gauge sectors in an
F-theory compactification; others can arise from D3-branes, which we
do not consider here.  The structure
of seven-branes is determined as follows.  
The elliptically fibered Calabi-Yau variety $X$ can be described in
Weierstrass form  \cite{Nakayama}
\begin{equation}
y^2 = x^3 + \, fx + \, g\,,
\label{eq:Weierstrass}
\end{equation}
where $f$ and $g$ are  functions that depend on
coordinates in $B$.  Specifically, $f$ and $g$ are sections of the
line bundles ${\cal O} (-4K_B)$ and ${\cal O} (-6K_B)$, where $K_B$ is
the canonical bundle of $B$.  
Seven-branes are located along the
discriminant locus $\Delta = 0$, where
\begin{equation}
\Delta = 4\, f^3 + 27 g^2.
\end{equation}
This is a complex codimension one locus in $B$ along which the
elliptic fiber becomes singular.  The possible singular fibers in
codimension one (originally analyzed for elliptic surfaces) have been
classified by Kodaira \cite{Kodaira}.

Up to an effect known as outer monodromy, which we will discuss later,
the gauge group is determined by the singular fibers over the
irreducible complex codimension one loci (divisors) $D$ that comprise
the discriminant locus; the individual gauge group factors can be
identified easily from the structure of $f$, $g$, and $\Delta$.
Specifically, if $f, g, \Delta$ vanish to specific orders on $D$,
\begin{equation}
  f \sim z^{{\rm ord}(f)}\qquad g \sim z^{{\rm ord}(g)} \qquad \Delta \sim z^{{\rm ord}(\Delta)},
\end{equation}
where $z$ is a local complex coordinate on $B$ with $D = \{z = 0\}$,
then the type of singular fiber above a generic point in $D$ can be
read off from Table \ref{table:Kodaira}; a seven-brane is located
along $z=0$.  In this paper we often use explicit coordinates such as
$z$, which can be thought of as local coordinates on a general complex
manifold, or homogeneous (Cox) coordinates when the manifold has a
toric description.  Divisors on which $f, g$ vanish to orders $(4, 6)$
are associated with non-minimal singularities that cannot be resolved
to a Calabi-Yau total space.  Such singular geometries lie at an
infinite distance from any fixed smooth geometry in the complex
structure moduli space, and we do not consider them here.  Codimension
two loci where $(f, g)$ vanish to degrees $(4, 6)$ correspond to
branches in the moduli space associated with tensionless string
transitions \cite{Seiberg-Witten, Morrison-Vafa-II}; these singular
Weierstrass models can be interpreted locally in terms of a
superconformal field theory (\cite{Seiberg-SCFT} see {\it e.g.}
\cite{Heckman-Morrison-Vafa, SCFT-2, SCFT-3, SCFT-4} for recent work);
a smooth Calabi-Yau resolution does not exist unless one first
performs a blow-up of the codimension two locus in the base. In this
paper we do not consider geometries with codimension two $(4, 6)$
loci.

Note from Table~\ref{table:Kodaira} that a distinguished role is
played by $SU(3)$ and $SU(2)$; unlike $SU(n>3)$, which can only be
realized by an $I_n$ fiber, $SU(3)$ and $SU(2)$ may also be realized
through a type $IV$ or type $III$ fiber.  One fundamental difference
between these singularity types is that the type $IV$ and $III$
singularity types are automatically imposed when $f, g$ vanish to
specified orders (as is also the case for  the $I_0^*, IV^{*}, III^{*}$
and $II^{*}$ types of singular fibers),
while the type $I_n$ singularities (like the type $I_{n>0}^*$
singularities) require the cancellation of nonzero terms in $f, g$ for
a higher degree of cancellation in $\Delta$.  This difference
underlies the distinction between those gauge groups that can be
realized in non-Higgsable clusters, and those that cannot.

\begin{table}[t]
\centering
\begin{tabular}{ccccc}
  \shs\shs Fiber Type \shs\shs & \shs{ {\rm ord}(f)}\shs & \shs{ {\rm ord} (g)}\shs & \shs{{\rm ord}($\Delta$)}\shs   &
   \shs Gauge Group \shs \\
  \hline \hline
  smooth & $\geq 0$ & $\geq 0$ & $0$ &  $-$\\
  $I_n$  & $0$ & $0$ & $n$ &   $SU(n)$ \\
  $II$ & $\geq 1$ & $1$ & $2$& $-$\\
  $III$ & $1$ & $\geq 2$ & $3$ &  $SU(2)$ \\
  $IV$  & $\geq 2$ & $2$ & $4$ &   $SU(3)$\\
  $I_n ^*$ & $2$ & $\geq 3$ & $n+6$ &  $SO(2n+4)$ \\
  $I_n ^*$ & $\geq 2$ & $3$ & $n+6$ &  $SO(2n+4)$ \\
  $IV^*$ & $\geq 3$ & $4$ & $8$ &  $E_6$ \\
  $III^*$  & $3$ & $\geq 5$ & $9$ &  $E_7$\\
  $II^*$  & $\geq 4$ & $5$ & $10$ &  $E_8$ \\ \hline \hline
\end{tabular}
\caption{Kodaira's classification of singular fibers, together with
  the associated gauge group (up to outer monodromy, which affects
  compactifications to 6D and 4D) in F-theory and
  degrees of vanishing along $f$, $g$, and $\Delta$.  Note that
  $SU(3)$ and $SU(2)$ are on a special footing compared to the other $SU(n)$
groups.
}
\label{table:Kodaira}
\end{table}

\sskip

A \emph{non-Higgsable cluster} arises when F-theory on $X$ exhibits a
gauge sector that cannot be spontaneously broken by a complex
structure deformation; that is, when the singularities associated to
the seven-branes exist at a generic point in the complex structure
moduli space of $X$, $CS(X)$. Mathematically, this occurs when a
connected set of complex codimension one loci (divisors) $D_i$ are
irreducible components of $f$ and $g$; that is, $f$ and $g$ always
vanish on $D_i$.  When ${\rm ord}_{D_i}(f) \geq 1, {\rm ord}_{D_i}(g)
\geq 2$, this forces a Kodaira singularity in $X$ associated with a
non-trivial gauge group.  This gauge group is, in fact, the gauge
group of an associated six-dimensional F-theory compactification for a
generic point in $CS(X)$; for this to be true in four dimensions, it
is also necessary to choose G-flux such that it does not affect the
gauge group.

The presence of non-Higgsable gauge groups can be seen easily in many
examples; if for the most generic $f$ and $g$ for a given base $B$ the
discriminant satisfies $\Delta\sim z^n$ for some $n>0$ and a local
coordinate $z$, then there is a seven-brane along $z=0$, and if $n >
2$ then the seven-brane configuration carries a non-Higgsable gauge
group.  Alternatively, if $\Delta \sim z^n$ with $n> 2$ only for a
specialization in the complex structure of $X$, {\it i.e.},  for non-generic
$f$ and $g$, then moving to a generic point in moduli space describes
spontaneous symmetry breaking via the Higgs mechanism.  Such a
symmetry breaking mechanism does not exist for non-Higgsable clusters.

It is straightforward to determine which individual gauge group
factors can possibly arise in a non-Higgsable cluster.  As discussed
above, these are precisely the gauge group factors associated with
Kodaira singularity types where the degree of vanishing of $\Delta$ is
forced by the degrees of vanishing of $f, g$.  This follows since in
general $f, g$ can be described as linear combinations of monomials
with free complex coefficients, and the only way a cancellation can be
arranged in $\Delta$ to a higher order than ${\text{min}}(3 \;
{\text{ord}(f)}, 2\; {\text{ord}(g)})$ is by choosing non-generic
values of the coefficients in $f, g$.  Thus, the gauge group factors
that can arise from a non-Higgsable structure are those associated
with all Kodaira singularity types except $I_n$ and $I_{n>0}^*$.

For example, consider the question of whether a type $IV$ fiber may be
non-Higgsable; assume that there exists a base $B$ and a local
coordinate $z$ such that for the most general $f$ and $g$ we have
\begin{equation}
  f=z^2\, \tilde f \qquad \text{and} \qquad g = z^2 \, \tilde g,
\end{equation}
{\it i.e.},  one can factor out an overall $z^2$ out of both $f$ and $g$.  If
this is the case, we see that $\Delta = z^4 \, (4\, z^2 \tilde f^3 +
27 \tilde g^2)$, and note that $\Delta$ automatically has the correct
order of vanishing for a type $IV$ fiber; {\it i.e.},  without requiring any
specialization in moduli space.  This shows that a type $IV$ fiber may in principle
arise as a non-Higgsable cluster.
Alternatively, note that
if one engineers a seven-brane with $SU(n)$ gauge symmetry arising
from an $I_n$ fiber, the order of vanishing of $f$ and $g$ along $z=0$
(${\rm ord}(f)={\rm ord}(g)=0$) does not ensure ${\rm ord}(\Delta)=n$; an additional
tuning is always needed, and thus an $I_n$ fiber cannot give rise to a
non-Higgsable cluster.

This simple analysis shows that fibers of type
$IV^*,III^*,II^*,IV,III,II,$ and $I_0^*$ may be realized as factors in a
non-Higgsable cluster.
There are no non-Higgsable clusters in eight dimensions, since in
eight dimensions all F-theory vacua are part of a single moduli space
where all nonabelian gauge group factors can be Higgsed.
To determine the complete set of gauge group factors that can be
associated with the
non-Higgsable Kodaira singularity types
for F-theory compactifications to 6D or 4D,
we must consider additional structure
known as {\it outer monodromy} that can act on a fiber in such a way that the
simply-laced gauge group factor becomes non-simply laced
 \cite{Bershadsky}.  
This occurs when a closed path on a divisor $D$ carrying a nontrivial
Kodaira singularity type brings the resolved fiber back to itself up
to a nontrivial permutation on its topological structure, which as
Kodaira showed is encoded by a set of rational curves ($\P^1$'s)
connected in the structure of the associated Dynkin diagram.
For example, there
is an action of $S_3$ on the Dynkin diagram of $SO(8)$; from a $\bZ_3$
subgroup of this action we see $SO(8)$ become $G_2$ as
\begin{equation}
\begin{tikzpicture}[scale=.4]
  \draw[thick] (-30: 3mm) -- (-30: 7mm);
  \draw[thick] (210: 3mm) -- (210: 7mm);
  \draw[thick] (90: 3mm) -- (90: 7mm);
  \draw[thick] (-30: 10mm) circle (3mm);
  \draw[thick] (210: 10mm) circle (3mm);
  \draw[thick] (90: 10mm) circle (3mm);
  \draw[thick] (0: 0mm) circle (3mm);
  \draw[thick,->] (60:10mm) arc (60:0:10mm); 
  \draw[thick,->] (-60:10mm) arc (-60:-120:10mm);
  \draw[thick,->] (-180:10mm) arc (-180:-240:10mm);
  \draw[xshift=3cm,->,thick] (0:0mm) -- (0:15mm);
  \draw[xshift=7cm,thick] (0,0) circle (3mm);
  \draw[xshift=7cm,thick] (0: 15mm) circle (3mm);
  \draw[xshift=7cm,thick] (0: 3mm) -- +(9mm,0);
  \draw[xshift=7cm,thick] (45: 3mm) -- +(10.6mm,0);
  \draw[xshift=7cm,thick] (-45: 3mm) -- +(10.6mm,0);
  \draw[xshift=7.85cm,thick] (120:0mm) -- (120:4mm);
  \draw[xshift=7.85cm,thick] (-120:0mm) -- (-120:4mm);
\end{tikzpicture}
\label{eqfig:D4G2Dynkin}
\end{equation}
where the nodes mapped to one another under the action are identified.
A geometry exhibiting this behavior gives rise to a $G_2$ gauge
theory. Similarly, outer monodromy can reduce the $SU(3)$ gauge group
on a type $IV$ singularity to $Sp(1)$ (which has the same Lie algebra as
$SU(2)$).  

\begin{table}[t]
\begin{tabular}{ccc}
  Fiber Type & \shs\shs Outer Monodromy \shs\shs & \shs\shs Gauge Group \shs\shs \\ \hline \hline
  $II^*$ & --- & $E_8$ \\
  $III^*$ & --- & $E_7$ \\ 
  $IV^*$ & --- & $E_6$ \\
  $IV^*$ & $\bZ_2$ & $F_4$ \\ 
  $I_0^*$ & --- & $SO(8)$ \\
  $I_0^*$ & $\bZ_2$ & $SO(7)$ \\
  $I_0^*$ & $S_3$ & $G_2$ \\
  $IV$ & --- & $SU(3)$ \\ 
  $IV$ & $\bZ_2$ & $Sp(1)\cong SU(2)$ \\ 
  $III$ & --- & $SU(2)$ \\
  $II$ & --- & --- \\ \hline \hline
\end{tabular}
\caption{Singular fibers in Kodaira's classification that
may give rise to non-Higgsable clusters, together with their possible
outer monodromies and associated gauge groups.}
\label{table:non-Higgsable classification}
\end{table}

Taking outer monodromy into account, the possible gauge groups that
can appear on a non-Higgsable seven-brane are given in Table
\ref{table:non-Higgsable classification}.  In each case, the existence
and type of outer monodromy can be determined by the structure of the
Weierstrass coefficients $f, g$ \cite{Bershadsky-all, Morrison-sn}.
In the case of interest to us here,  
for a type $IV$ singularity the gauge group is $SU(2)$ unless the
leading coefficient in the expansion of $g$ in a local coordinate $z$
around the locus in the base supporting the singularity is a perfect
square, {\it i.e.}, $g = g_2z^2 + g_3z^3+ \cdots = z^2 \gamma^2 + {\cal O}
(z^{3})$, where $\gamma$ can be an algebraic function of the remaining
coordinates.  For a type $IV$ singularity in a non-Higgsable cluster
to give rise to a non-Higgsable $SU(3)$ gauge group, the coefficient $g_2$ must be
a single (even) monomial or a constant, or it is not generically a
perfect square.  We will present examples of this type.

While the preceding discussion describes individual gauge factors in a
non-Higgsable cluster, it is also possible for non-Higgsable clusters
to contain multiple gauge group factors as well as matter charged
under these gauge group factors.  For six dimensions, the set of
possible non-Higgsable clusters of gauge groups and associated minimal
matter content for F-theory compactifications on Calabi-Yau threefolds
over base surfaces $B_2$ were classified in \cite{clusters}.  As
mentioned above, these include clusters that give rise to $G_2 \times
SU(2)$ and $SU(2) \times SO(7) \times SU(2)$ gauge groups, with matter
that is jointly charged under each adjacent pair of gauge groups in
the cluster.  A similar set of clusters are present at the geometric
level for base threefolds.  For 4D F-theory compactifications with
smooth heterotic duals, the only possible non-Higgsable clusters are
those with only a single gauge group factor (but include $SU(2)$,
which cannot be realized by itself as a non-Higgsable gauge group in
6D compactifications); an analysis of the non-Higgsable structures in
F-theory models with smooth heterotic duals was carried out in
\cite{Anderson-WT}.  The range of product groups that can be
realized in non-Higgsable clusters also increases substantially for
four dimensional F-theory compactifications \cite{mt-4D-clusters}.  In
particular, as we demonstrate with some specific examples in this
paper, it is possible in a 4D F-theory compactification to have a
non-Higgsable cluster with the gauge group structure $SU(3) \times
SU(2)$.

\sskip

The discussion so far has focused on the gauge group
of an F-theory compactification.
Of course, for a complete specification of the physics of a low-energy
theory arising from F-theory we also need to describe the matter
spectrum of charged particle states that arise in the theory, and the
associated representation theory.  

A geometric description of the physics of general seven-brane
configurations that gives substantial insight into the nonperturbative
structure of these objects is given using the particle states that end
on seven-branes in F-theory.  These states can be described using the
language of $(p,q)$ string junctions
\cite{Schwarz-SJ,Gaberdiel-Zwiebach,DeWolfe-Zwiebach}.  $(p, q)$ strings are bound
states of $p$ F-strings and $q$ D-strings; these are the
F-theory limit of particular M2-brane configurations in a defining
M-theory compactification, namely those where the M2-brane wraps the
one-cycle $p\, \pi_1 + q\, \pi_2$ in the elliptic fiber, as well as
extending in $B$.  Upon movement in the complex structure moduli space
of $X$, these strings, and more complicated branched
\emph{string junctions} formed from them,
 can shrink to zero size, in which case they
become massless.  If this occurs along a codimension one locus in $B$,
the string junctions describe vector multiplets; if along a
codimension two locus, they live in matter multiplets ({\it i.e.},
hypermultiplets and chiral/anti-chiral  multiplets in six- and four-dimensional
compactifications, respectively).  For recent studies of the connection
between deformation theory and string junctions, see
\cite{GHS-I,GHS-II}.

The representation theoretic matter content can also be read off by
performing a K\" ahler resolution in M-theory \cite{Bershadsky-all,
  Katz-Vafa}.  This type of resolution of singularities has been used
to analyze the representation theory structure of matter in a wide
variety of geometries.
Despite a substantial amount of work on this topic
({\it e.g.} \cite{GrassiMorrisonFirst, mt-singularities,
  Esole-Yau, GrassiMorrison, Lawrie-Sakura, Hayashi-lms, Esole-sy}), there is
still no complete classification of the types of codimension two
singularities that can arise in F-theory, and no complete dictionary
that relates such codimension two singularities to arbitrary
representations of the possible gauge groups.  For the purposes of
this paper, however, we will only need a simple set of
representations, namely matter states that transform in the
fundamental representation of a single $SU(n)$ gauge factor, or
bifundamental representations that transform in the fundamental (or
anti-fundamental) of each of a pair of product gauge factors $SU(n)
\times SU(m)$.

One important point that we emphasize again here is that while in six
dimensions the gauge group and matter content encoded in the geometry
of an F-theory compactification correspond precisely to the physical
gauge group and matter content, in four dimensions the connection is
much less direct.  Additional features such G-flux, D3-branes, and
additional moduli are present in 4D F-theory compactifications, and
can modify both the gauge group and matter content of the physical
theory from those which are seen in the geometric description in terms
of the Calabi-Yau fourfold.  In particular, a detailed determination
of the matter content of a 4D F-theory compactification depends upon
consideration of these factors.  In this paper we primarily analyze
the geometric structure of compactifications, and assume that in a
broad class of cases the physical structure of the low-energy theory
will match the geometric structure, as often assumed in the F-theory GUT
literature.  We address some of the issues involved in connecting the
geometric picture to the low-energy physics and determining precise
features of the matter spectrum in sections \ref{sec:issues} and
\ref{sec:discussion}.

\section{Constructing the Standard Model Spectrum in F-theory}
\label{sec:SM}

In this section we describe in very general terms the structure of any
4D F-theory model that contains the gauge group $G_{321} =SU(3) \times
SU(2)\times U(1)$ at the level of geometry; in particular, we allow
for the use of non-Higgsable clusters.  The basic approach taken is
based on the geometry of the base $B_3$ of the elliptically fibered
Calabi-Yau fourfold $X, \pi: X \rightarrow B_3, \pi^{-1} (p) \cong \E$
$\forall p \in B_3 \setminus \Delta$.  By focusing on the structure of
the base $B_3$ many aspects of the structure and classification of
possible models are simplified.  In particular, for any given base
$B_3$ there can be many different elliptically fibered Calabi-Yau
varieties with different singularity structures associated with
different gauge groups, realized by tuning to substrata in complex
structure moduli, {\it i.e.}  to a specialized Weierstrass model.  By
grouping these together and focusing on the physics of the generic
elliptic fibration over any given base, the classification problem is
substantially simplified.

In broad strokes, any 4D F-theory model that contains the gauge group
$G_{321}$ can be constructed and described in the following steps:
\vspace*{0.1in}

\begin{itemize}
\item[A)] {\bf Choose a base $B_3$.}
A threefold must be chosen that can support an elliptically fibered
fourfold.  Many examples of such threefolds are known, though the
complete set of possibilities has not been classified.
\vspace*{0.1in}

\item[B)] {\bf Check for non-Higgsable clusters.}
From the most general Weierstrass model over a given base $B_3$ it is
possible to determine whether or not the geometry exhibits a
non-Higgsable cluster.  In particular, we can
check to see if there are
$SU(3)$ and/or $SU(2)$ non-Higgsable clusters.  
We utilize several different methods in this paper to check for the
existence of non-Higgsable clusters in 4D F-theory compactifications.
  \vspace*{0.1in}
\item[C)]{\bf Tune non-abelian factors as necessary.}  If the
  non-Higgsable part of the gauge group does not contain $SU(3) \times
  SU(2)$, the rest of the nonabelian part of $G_{321}$ can be tuned by
  going to a special locus\footnote{Note that this may not be
    possible in some bases. For example, there are bases $B_3$ that
    contain a non-Higgsable $SU(3)$, where there is no possible tuning
    of a Weierstrass model having an additional $SU(2)$ on an
    intersecting divisor without producing a $(4, 6)$ singularity at a
    codimension two locus on the base.} in the moduli parameterizing the generic
  Weierstrass model over $B_3$.   \vspace*{0.1in}

\item[D)]{\bf Identify a configuration with a $U(1)$ factor.}  F-theory models
  with abelian $U(1)$ factors correspond to Calabi-Yau manifolds with
  nontrivial rational sections that live in a Mordell-Weil group of
  nonzero rank.   Identifying the Mordell-Weil group of a
  Weierstrass model over a given base is generally a difficult
  mathematical problem, but methods exist for constructing general
  models with a single $U(1)$ factor over any given base.
\end{itemize}

Of course, obtaining
$G_{321}$ is necessary but not sufficient to realize a model that
contains the standard model spectrum; the geometry must exhibit
particular representations of this group in order to match the field
content of the SM.  In the MSSM, for example, the matter fields are
chiral superfields in  representations of $G_{321}$
\begin{center}
\begin{tabular}{lll}
  $Q: (3,2)_{1}$ &
  $\qquad U: (\ov 3,1)_{-4}$ &
  $\qquad D: (\ov 3,1)_{2}$  \\
  $L: (1,2)_{-3}$ &
  $\qquad E: (1,1)_{6}$ &
  $\qquad N: (1,1)_0$  \\
  $H_u: (1,2)_{3}$ &
  $\qquad H_d: (1,2)_{-3}$ &
\end{tabular}
\end{center}
and obtaining a realistic theory requires identifying the fields $Q,
U, D, L, E$, at the very least, possibly augmented by $N, H_u,$ and
$H_d$ and/or other representations depending on the model.
Identification of the representations of $G_{321}$ that appear in the
geometry can be done following step (D), by analyzing codimension two
singularities using deformation or resolution methods.  As discussed
earlier, we focus here purely on the
geometric analysis of the gauge group and matter content. A more
complete analysis would need to incorporate G-flux and other features
that might affect the low-energy  matter spectrum; in
particular, though the correct Lie algebra representations of the
standard model may emerge geometrically, obtaining a chiral spectrum
requires the introduction of G-flux.  

For models with a suitable spectrum, in principle more detailed
aspects of the standard model could be checked.  In this paper we
primarily focus on the construction of models that have the nonabelian
$SU(3) \times SU(2)$ structure of the standard model, though we also
carry out a limited analysis of $U(1)$ factors and matter spectra both
in general and in specific cases.

\sskip

In principle, this approach could be used to systematically identify
large classes of F-theory models that contain the gauge group
$G_{321}$ that appears in the standard model of particle physics.
More generally, this approach could be used to systematically describe
elliptically fibered Calabi-Yau manifolds that give F-theory models
with any gauge group.  A more detailed description of how this general
approach can in principle be used to describe all elliptically fibered
Calabi-Yau \emph{threefolds}, and some technical challenges to a complete
classification using this approach are described in
\cite{Johnson-Taylor}.  In addition to the issues described in that
paper, for fourfolds there is a further complication in giving a complete
mathematical characterization of possible bases $B_3$ that support
elliptically fibered Calabi-Yau fourfolds, which requires more
sophisticated mathematics than the classification of bases $B_2$ for
threefolds.  Notwithstanding the challenges of finding a complete
classification of fourfolds for realistic F-theory models, this
approach can give a broad class of models with semi-realistic
phenomenological features, and all F-theory constructions of physical
theories with standard model-like features must be describable in this
general framework.

We now describe some more detailed aspects of each of the steps above
in turn.

\subsection{Choose a base $B_3$}

A broad class of bases $B_3$ that can support 4D F-theory models are
known.  Many Calabi-Yau fourfolds that arise as hypersurfaces or
complete intersections of toric varieties have been studied using the
original approach of Batyrev \cite{Batyrev}
(see for example \cite{Kreuzer-Skarke-4D, lsw-toric}); many of these fourfolds
have a description as an elliptic fibration over an appropriate toric
base $B_3$ \cite{Candelas-cs}
and are appropriate for F-theory compactifications
\cite{Klemm-4D, Berglund-Mayr, Mohri, Grassi-network, Grimm-Taylor}.  A complete classification of all bases $B_3$ that have
the form of $\P^1$ bundles over a complex surface $B_2$ and that
support elliptically fibered Calabi-Yau fourfolds giving
F-theory models with smooth heterotic dual constructions was given in
\cite{Anderson-WT}; in these cases the base $B_2$ is a generalized del
Pezzo surface.  A much broader class of bases $B_3$ can
be constructed as $\P^1$ bundles over surfaces $B_2$ that can act as
bases for elliptically fibered threefolds; a complete list of all of
the over 100,000 toric and ``semi-toric'' surfaces $B_2$ of this type
has been constructed \cite{toric, Martini-WT}, and a systematic
analysis of $\P^1$ bundles over such bases will appear elsewhere
\cite{Halverson-WT}.  Most of the explicit examples considered later
in this paper use bases $B_3$ of this form.  Recent constructions of
Calabi-Yau fourfolds as complete intersections, of which more than
99.9\% satisfy appropriate conditions for elliptic fibration structure
\cite{Gray-hl, Gray-hl2}, also promise to provide a rich supply of
examples.

Following the
framework of Mori theory \cite{Mori}, an even broader class of base
threefolds $B_3$ can in principle be constructed, though unlike the
case of Calabi-Yau threefolds, where all minimal bases are known
\cite{Grassi}, a complete set of minimal threefold bases $B_3$ from
which all others can be constructed by suitable geometric transitions
is not yet known.  For Calabi-Yau threefolds, the set of allowed bases
$B_2$ is connected by blowing up and down points on the base,
corresponding to tensionless string transitions \cite{Seiberg-Witten,
  Morrison-Vafa-II} in the associated F-theory models.  While the
bases $B_3$ for elliptically fibered Calabi-Yau fourfolds are
similarly connected through birational (blow-up and blow-down)
transitions, it is
not known if the total space of $B_3$'s has only one or more connected
components.

\subsection{Check for non-Higgsable clusters.}
\label{sec:check-clusters}

Having chosen a base $B_3$, the next step is to consider the general
Weierstrass model over that base, and determine whether or not there
are non-Higgsable clusters.  There are a number of ways of doing this,
depending on the geometry of the base.  When the base is toric, this
computation is straightforward.  Toric bases can be described either
using a gauged linear
sigma model language common in physics (see {\it e.g.}
\cite{Knapp-Kreuzer}), or equivalently
in terms of a toric fan as is standard in
mathematics \cite{Fulton}; in toric cases there is a straightforward
algorithm for determining all monomials in the Weierstrass
coefficients $f, g$ through solutions to a given set of inequalities,
and one can check directly the order of vanishing on each of the toric
divisors, as described for example in \cite{toric}.  

There are also systematic methods that can be applied when the base is
not toric.  When the base is a surface $B_2$, the divisor class $[-
K]$ can be formally decomposed over $\Q$ into irreducible components;
this {\it Zariski decomposition} can be carried out using the
intersection properties of curves on the surface, and determines the
minimal degrees of vanishing of $f, g$ on any curves in the base.
This was the approach used in \cite{clusters} to identify all
non-Higgsable clusters in six dimensions.  When the base is a general
non-toric threefold $B_3$ the analogue of the Zariski decomposition is
somewhat more complicated to describe in terms of the intersection
numbers on the threefold, but a related method of analysis in terms of
the geometry of divisors and curves can determine the presence of
non-Higgsable clusters.  An explicit way to determine minimal
vanishing degrees of $f, g$ on a divisor in a general base threefold
using information about the line bundles over the divisor that contain
the coefficients in an expansion of $f, g$ around that divisor is
developed in \cite{Anderson-WT, mt-4D-clusters}.  

For concreteness, we present here a simple example that demonstrates
the appearance of a non-Higgsable cluster using the gauged linear
sigma model approach
to toric geometry that may be most familiar to physicists.  For the
examples described later in the paper, we primarily use the fan
description of toric varieties and/or the more abstract description of
the line bundles in which the Weierstrass coefficients take values.
All of these approaches can be used in more complicated examples and
may be easily analyzed on a computer.

Consider the possibility
\begin{equation}
  \label{eq:NHCexbase}
  B_3 = \bP^1 \times \bF_{8}
\end{equation}
where $\bF_{8}$ is the eighth of the infinite series of Hirzebruch
surfaces $\bF_n$ that themselves describe $\P^1$
bundles over $\P^1$.  This base $B_3$ is a toric variety and can be
described as a quotient space using the following coordinates and $\bC^*$ actions
\begin{center}
  \begin{tabular}{c|cccccc}
    & $\,\,x_1\,\,$ & $\,\,x_2\,\,$ & $\,\,x_3\,\,$ & $\,\,x_4\,\,$ & $\,\,x_5\,\,$ & $\,\,x_6\,\,$ \\ \hline \hline
    $\bC_1{^*}$ & $1$ & $1$ & & & &  \\ \hline
    $\bC_2{^*}$ & & &$1$ & &$1$ &$ 8$  \\ \hline
    $\bC_3{^*}$ & & & &$1$ & & $1$ \\ \hline
  \end{tabular}
\end{center}
where empty entries are zero.  We also define the set
\begin{equation}
S = \{x_1=x_2=0\} \cup \{x_3=x_5=0\} \cup \{x_4=x_6=0\}.  
\end{equation} 
Then $B_3$ is defined by taking the $(x_1,\dots,x_6)\in \bC^6$,
removing $S$, and quotienting by the $\bC^*$ actions
\begin{equation}
B_3 = \frac{\bC^6 \setminus S}{\bC_1^* \times \bC_2^* \times \bC_3^* }\,,
\end{equation}
where the $\bC^*$ actions give equivalence classes; for example, the $\bC_2^*$
quotient identifies
\begin{equation}
(x_1,\dots,x_6) \sim (x_1,x_2,\lambda\, x_3,x_4,\lambda x_5, \lambda^{8} x_6) \,\,\, \lambda \in \bC^*
\end{equation}
and similarly for the other actions.  In the quotient space, $(x_1,\dots, x_6)$
are homogeneous coordinates.
Such a quotient space can arise as the set of supersymmetric ground
states of an appropriate gauged linear sigma model.

Having defined $B_3$ in this example, we may now construct the
most general  Weierstrass model (\ref{eq:Weierstrass}).  Here
$\cO(-K_B) = \cO(2,10,2)$,
meaning that sections of the line bundle associated with the divisor
$-K_B$  transform with the given powers of the transformation
parameters under the three $\C^*$ actions; this result is obtained by taking the sum of the charges
of the coordinates under the $\bC^*$ actions.  It follows that
$f$ and $g$ are global sections of the line bundles
\begin{equation}
f \in \cO(8,40,8) \qquad \qquad g \in \cO(12,60,12).
\end{equation}
Each monomial in $f$ therefore scales with
degrees $8$, $40$, and $8$ under $\bC_1^*$, $\bC_2^*$, and $\bC_3^*$,
respectively,  and similarly for monomials in $g$.  Furthermore, the
exponent of each monomial must be
non-negative.  The monomial
\begin{equation}
x_1^3\,\, x_2^5 \,\, x_3^5\,\, x_4^4\,\, x_5^3\,\, x_6^4,
\end{equation}
for example, scales as a
monomial in $f$.  It is a simple exercise to determine all such allowed
monomials in $f$, and thereby to determine the most general $f$.  A
similar computation with different scaling degrees holds for $g$.

Note the following interesting feature, however: $f$ may not
contain a monomial $x_6^{n}$ for $n > 5$ since it would oversaturate
the scaling degree of $\bC_2^*$ and the powers of $x_i$ 
(in particular $x_4$)
in any monomial
must be non-negative.  Thus $n$ must  satisfy $n\le 5$ and therefore
every monomial in $f$ must have $x_4^m$ with $m\ge 3$ in order to saturate
the scaling degree of $\bC_3^*$.  A similar analysis for $g$ shows that 
the monomials in $f, g$ can always be written in the form
\begin{equation}
f = x_4^3\,\, \tilde f \qquad \qquad g = x_4^5\,\, \tilde g.
\end{equation}
We see that the most general $f$ and $g$ have overall factors
of $x_4$, where $x_4 = 0$
defines a divisor in $B_3$.  This is the characteristic signature of a non-Higgsable
cluster, and from table \ref{table:Kodaira} we identify that there
is an $E_7$ non-Higgsable cluster along $x_4=0$.  Note that for simplicity
we chose an example where outer monodromy would not be in effect.

\subsection{Ensure the existence of $SU(3) \times SU(2)$}

At this point in the systematic process we are describing we have
chosen a base $B_3$ and have identified all non-Higgsable clusters
exhibited by the generic Weierstrass model over $B_3$.  If there are
$SU(3)$ or $SU(2)$ non-Higgsable gauge group factors we do not have to
specialize in moduli space to ensure their existence.

We would now like to discuss broadly the possible ways to realize the
non-abelian sector of the standard model.  Specifically, if the gauge
group of the four-dimensional compactification is to contain the
subgroup $G_{32}= SU(3) \times SU(2)$, this group must either arise
from a non-Higgsable cluster, or the Weierstrass model must be tuned
to realize whatever part of this group is not found through a
non-Higgsable cluster.  Furthermore, for quark-like matter to arise
that is charged under both factors, the divisors supporting these two
gauge group factors must intersect in a curve in $B_3$.  There are
several logical possibilities:

\begin{enumerate}
\item[i)] {\bf No non-Higgsable gauge group.}  In this case the
  Weierstrass model must either be tuned to contain both the $SU(3)$
  and $SU(2)$ gauge group factors on intersecting divisors, or
    to have have a grand unified gauge factor on a single non-rigid
divisor
that contains $SU(3) \times SU(2)$
as a subgroup.
\item[ii)]
{\bf Non-Higgsable $SU(2)$.}
In this case either an additional $SU(3)$ must be tuned or the
original $SU(2)$ must be enhanced to an $SU(3)$ and an additional
$SU(2)$ tuned on an intersecting divisor.
\item[iii)]
{\bf Non-Higgsable $SU(3)$.}
An additional $SU(2)$ must be tuned on a divisor that intersects the
divisor carrying the $SU(3)$.
\item[iv)]
{\bf Non-Higgsable $SU(2) \times SU(2)$.}
If there is a non-Higgsable $SU(2) \times SU(2)$ on two intersecting
divisors, the Weierstrass model must be tuned so that one of
them is enhanced to $SU(3)$.
\item[v)]
{\bf Non-Higgsable $SU(3) \times SU(2)$.}
In this case the entire desired nonabelian gauge group is
automatically present in the generic Weierstrass model over $B_3$ and
no further tuning is necessary.
\end{enumerate}

\noindent In almost all cases considered previously in the literature,
the approach taken is that of case i).  This is the only approach
possible when the gauge group factors $SU(3) \times SU(2)$ can be
realized geometrically in an F-theory construction within a unifying
group $SU(5)$, so that the $SU(3)$ and $SU(2)$ factors are realized on
divisors in the same homology class, since $SU(5)$ is associated with
a type $I_5$ singularity, which cannot arise on any divisor supporting
a non-Higgsable cluster.  For $SO(10)$ and $E_6$ unification
scenarios, the set of possibilities is somewhat richer.  $E_6$ can be
realized through a non-Higgsable cluster, and could then be broken
down to a standard model gauge group through a flux on the seven-brane
world volume.  While $SO(10)$ cannot be realized through a
non-Higgsable cluster, a divisor with a non-Higgsable type $I_0^*$
singularity (or a type $III$ or $IV$ singularity) could be enhanced to
$SO(10)$, and in principle this group could be broken down to $SU(3)
\times SU(2)$ in such a way that part of the gauge group was still
non-Higgsable.  Approaches to GUT constructions using approach i) have
been extensively studied in the literature, beginning with
\cite{Donagi-Wijnholt, Beasley-hv, Beasley-hv2}, and extending to
global constructions
\cite{Marsano-F-theory,Marsano-flux,Blumenhagen-F-theory,Cvetic-gh,Knapp:2011wk};
for reviews see \cite{Heckman-review, Weigand-review}.  In much of
this work, internal flux on the seven-branes is the mechanism used for
GUT breaking, so while these investigations have mostly focused on
constructions of type i), an extension to include non-Higgsable
structures within $SO(10)$ or $E_6$ models may be natural.  In some
recent work, such as \cite{LinWeigand, AllToricHypFibs}, approach i)
has been taken but without the GUT assumption, so that the divisors
supporting the $SU(3)$ and $SU(2)$ factors are assumed to be distinct
and intersecting.  These constructions are closer in spirit to those
we consider in the rest of this paper; the difference is that the
constructions we focus on here correspond to cases ii)-v).

In all cases
other than i) above, at least one of the nonabelian gauge group
factors is realized through a non-Higgsable cluster.  One of the
primary points of this paper is that the other possibilities can be
realized naturally in F-theory and offer some interesting
phenomenological features.  We focus in particular on the cases iii)
and v), where the $SU(3)$ nonabelian factor is non-Higgsable,
motivated by the observed fact that the $SU(3)$ of QCD observed in
nature is unbroken.  As we show in \S\ref{sec:examples}, such
non-Higgsable clusters can arise in specific simple examples of base
threefolds $B_3$.  Note that while a single non-Higgsable
$SU(3)$ or $SU(2)$ factor can easily arise in an F-theory model with a
smooth dual heterotic description, if there is a non-Higgsable $SU(3)$
and $SU(2)$ is realized on a separate intersecting divisor, this would
correspond to a singular geometry in any dual heterotic description.
Thus, the cases iii) and v), of a non-Higgsable QCD $SU(3)$ group, are
most clearly visible from geometry in the F-theory approach.

The classification of $SU(3)$ and $SU(2)$ factors above is according
to whether or not they arise from non-Higgsable groups, but for each
of these gauge group factors, a more refined set of cases can be
distinguished based on the precise Kodaira singularity types that
realize each factor.  As discussed earlier, both $SU(3)$ and $SU(2)$
can be realized in different ways through different Kodaira
singularities that involve different numbers of seven-branes.  A
detailed list of the possibilities is given in
Table~\ref{t:3-2-realizations}.  Some further comments on the table
may be relevant.  First, only the $I_3$ and $I_2$ realizations exist
at weak coupling and these cannot be non-Higgsable; the states
realizing the others are non-trivial $(p,q)$ string junctions
associated with type $III$ or type $IV$ singularities, which can
either be Higgsable or non-Higgsable.  Second, the $SU(2)$ realization
involving a type $IV$ fiber is an $Sp(1)\cong SU(2)$ realization where
there is outer monodromy on the type $IV$ fiber.

\begin{table}[t]
  \centering
  \begin{tabular}{cccc}
    & $\exists$ Higgsable  & $\,\,\exists$ Non-Higgsable  & $\,\,$ $\#$ Branes$\,\,$ \\ \hline \hline
    $SU(3)$ from $I_3$ & Yes & No & $3$ \\
    $SU(3)$ from $IV$ & Yes & Yes & $4$ \\ \hline
    $SU(2)$ from $I_2$ & Yes & No & $2$ \\
    $SU(2)$ from $III$ & Yes & Yes & $3$ \\
    $SU(2)$ from $IV$ & Yes & Yes & $4$ \\ \hline \hline
  \end{tabular}
  \caption{  
    Distinct
    realizations of $SU(3)$ and $SU(2)$ in
    F-theory through different Kodaira singularity types, together with the number of seven-branes realizing the
    gauge theory and whether or not there exist Higgsable and non-Higgsable
    configurations.}
  \label{t:3-2-realizations}
\end{table}

The case v) is associated with a non-Higgsable cluster that arises
when there are forced type III and IV Kodaira type singularities on
intersecting divisors, or two intersecting type IV singularities where
one has monodromy giving an $SU(2)$.  A particularly interesting
aspect of the first of these geometries is that the minimal $(1, 2,
3)$ and $(2, 2, 4)$ vanishing degrees of $f, g, \Delta$ on the
divisors carrying the gauge group factors force vanishing degrees of
at least $(3, 4, 8)$ on the intersection curve.  Here the number of
seven-branes, associated with the degree of vanishing of $\Delta$, is
increased to 8 rather than $3+4 = 7$, since ord $(\Delta)\geq {\rm
  max}( 3\,{\rm ord}(f), 2\, {\rm ord}(g))$.  With a 
generalization of the results of \cite{Katz-Vafa}, as we describe
below, the resulting matter is that associated with the embedding of
$SU(3) \times SU(2)$ in the adjoint of $E_6$, with interesting phenomenological
properties.  

Note that in all cases that we consider here, the intersection between
the divisors supporting the $SU(3)$ and $SU(2)$ factors is assumed to
be transverse.  Other constructions, for example where the divisors
are tangent at the intersection point, could also be investigated.

\subsection{Identify a configuration with a $U(1)$ factor.}

We have identified the possible ways to engineer $SU(3)\times SU(2)$
in F-theory, but have not yet addressed the $U(1)$ factor in
$G_{321}$.  If $G_{32}$ arises as a subgroup of some grand unified
group $G_{GUT}$, the weak hypercharge $U(1)_Y$ may arise from the the
non-abelian structure of $G_{GUT}$.  If not, $U(1)_Y$ must arise from
a non-trivial Mordell-Weil group of sections of the elliptic
fibration.  The number of $U(1)$ factors is given by the rank of the
Mordell-Weil group \cite{Morrison-Vafa-II}, and a realistic theory
requires that one may be identified as the weak hypercharge.  

In general, the problem of identifying the Mordell-Weil group is a
difficult mathematical problem.  Much recent work has focused on
F-theory models with one or more $U(1)$ factors, and while there is
some understanding of models with small Mordell-Weil rank, a general
understanding of the geometry of models with generic Mordell-Weil rank
is still lacking.  In the case of a single $U(1)$ factor, however,
there is a relatively clear understanding of how models can be
constructed.  The generic form of a Weierstrass model for a Calabi-Yau
elliptic fibration with a section corresponding to Mordell-Weil rank
one or higher is given in \cite{Morrison-Park}.  We use this approach
in Section \ref{sec:abelian} to describe some aspects of the tuning of
an additional $U(1)$ factor in models with a non-Higgsable $SU(3)
\times SU(2)$ gauge group. 
In the recent works \cite{LinWeigand,AllToricHypFibs}
progress was made in realizing the weak hypercharge in F-theory
by engineering a non-trivial Mordell-Weil group, though as mentioned
above
those analyses
did not utilize non-Higgsable clusters. 

\subsection{Low-energy physics, matter, and G-flux} 

As mentioned at the beginning of this section, the preceding
discussion is carried out at the level of pure geometry.  While in
six-dimensional theories the spectrum of massless states in a
low-energy supergravity theory is simply related to the geometric
structure of the associated Calabi-Yau threefold, for 4D theories
there are additional effects which may or may not modify the geometric
gauge group.  G-flux and possible other effects such as D3-branes can
modify both the geometric gauge group and matter content of the
theory.

On the one hand, G-flux can correspond to flux in the world-volume of
a seven-brane, which can break a geometrically non-Higgsable gauge
group factor.  On the other hand, the superpotential induced by G-flux
can push the theory to a special sublocus of moduli space where the
symmetry is enhanced beyond the geometric gauge group obtained a
generic point in complex structure moduli space.  G-flux can also
affect the matter content of the theory.  The geometric analysis
captures the Lie algebraic structure of possible matter states, but
only describes non-chiral $\cN = 2$ type matter.  G-flux can modify
the matter spectrum, producing chiral matter fields in the various
representations identified through geometry.  While in this paper we
focus on the geometric structures involved, in section
\ref{sec:minimal chiral} we utilize an anomaly analysis to determine
the minimal chiral $SU(3)\times SU(2)\times U(1)$ which may be
realized in the non-Higgsable QCD scenario we propose. In general,
though, a full understanding of any realistic realization of the
standard model spectrum in F-theory will require a detailed analysis
of the role of G-flux, the superpotential, and the chiral spectrum.
We describe some further details of the issues involved in
\S\ref{sec:issues-G-flux}.

\section{Non-Higgsable QCD}
\label{sec:nonhiggsableQCD}

In the last sections we have studied the possible ways to realize the
standard model in F-theory. We have also seen that F-theory exhibits
special realizations of $SU(3)$ and $SU(2)$ gauge theories on
seven-branes that do not exist for other $SU(n)$, and furthermore that
these realizations are the only $SU(n)$ realizations in F-theory that
may be non-Higgsable. Given these facts and the existence of an
unbroken QCD sector in nature, in this and the following sections we
will focus in particular on the possibility of realizing $SU(3)_{QCD}$
via a non-Higgsable seven-brane.

To consider such a possibility, we will need to recall the possible
realizations of $SU(n)$ gauge theories in F-theory.
\begin{itemize}
\item[1.] $SU(n)$ may be realized by an $I_n$ fiber for any $n$,
  but these are necessarily Higgsable and are the F-theory lift of
  $n$ coincident $D7$-branes.
\item[2.] $SU(3)$ may also be realized via a type $IV$ fiber without
  outer monodromy. This configuration utilizes \emph{four}
  seven-branes and does not admit a weakly coupled type IIB
  description, as it necessarily involves non-trivial $(p,q)$ string
  junctions. Such a configuration could potentially be non-Higgsable, but
  is not required to be.
\item[3.] $SU(2)$ may also be realized via a type $III$ fiber without
  outer monodromy, or a type $IV$ fiber with outer monodromy. Both could
  potentially be non-Higgsable, but are not required to be. 
\end{itemize}
We find it suggestive that $SU(3)$ and $SU(2)$ are the only gauge factors
that may be non-Higgsable.

Given these facts, we see there are three qualitatively different ways
to realize non-Higgsable QCD.  This follows for a simple reason: there
is one possible realization of a non-Higgsable $SU(3)$, namely from a
seven-brane associated to a type $IV$ singular fiber, but it may be
paired with any of three possible realizations of $SU(2)$. These
possibilities are realized by seven-branes associated to a type $IV$
fiber with outer monodromy, a type $III$ fiber, or an $I_2$ fiber. We
therefore refer to these realizations of non-Higgsable QCD as
$IV$-$IVm$, $IV$-$III$, and $IV$-$I_2$, respectively, according to
their realization of the standard model gauge subgroup $SU(3)\times
SU(2)$. We will focus slightly more on the $IV$-$III$ realization
since the spectrum at the $IV$-$III$ intersection is most interesting.

We emphasize from the outset that in this classification
of the three realizations of non-Higgsable QCD, we have not made any
assumptions as to whether or not the $SU(2)$
factor is also non-Higgsable. Though
the $IV$-$I_2$ case always has a Higgsable $SU(2)$, we will see in
examples that either a Higgsable or non-Higgsable $SU(2)$ are possible
in the $IV$-$III$ and $IV$-$IVm$ cases.

\sskip

In studying non-Higgsable QCD in F-theory models, we find it helpful to use
a variety of techniques, in part because each has different advantages.  In this
section we review the string junction description of $SU(3)$ and
$SU(2)$ seven-branes realized by type $IV$ and $III$ fibers,
respectively. 
We also present the first junction analysis of the
geometry near a $IV$-$III$ collision and discuss the implications for
matter spectra. 
One feature of the junction analysis it that it
demonstrates the possible existence of $SU(3)\times SU(2)$ singlet
states in the $IV$-$III$  realization, which may be interpreted as
right-handed electrons or neutrinos in some models.

In analyzing the F-theory geometry of the various non-Higgsable QCD
models we generally begin by studying the gauge group and then
consider the matter content at the level of Lie algebra representations
realized in the geometry.  The different constructions of the gauge
group have already been reviewed in the previous sections, and the
junction analysis gives a deeper perspective on the role of
seven-branes in these different geometries.
  
The analysis of the matter content is somewhat more complicated.
There are a number of ways to analyze matter representations
associated with codimension two singularities in F-theory.  The
original Katz-Vafa analysis \cite{Katz-Vafa} of the simplest types of
codimension two singularities --- namely those where the Kodaira type
of the codimension two singularity has a rank one greater than that of
the generic point in the codimension one locus --- involves a
straightforward decomposition of the adjoint representation of the
larger group; even there, however, some care must be taken in
determining the multiplicity of each representation.  For 6D theories,
anomaly cancellation \cite{gswest, Sagnotti, Sadov, Erler} provides a
powerful tool that in many cases uniquely determines the
representation content of a given theory, including multiplicities
\cite{GrassiMorrisonFirst, KMT-II, GrassiMorrison, kpt,
  Johnson-Taylor}, though this approach does not detect singlet
representations of the nonabelian gauge group that can play an
important role, for example, in realizations of the standard model.
As discussed in section \ref{sec:F-theory-clusters}, for 4D theories
with more complicated singularities a resolution of the singularity
can be achieved by standard methods in algebraic geometry, and more
recently the method of deformation has been developed as a useful
alternative to analyzing such singularities.

We focus in this section on the deformation method using junctions and
in the next section on 6D
anomaly analyses.  Since a number of factors are relevant in
determining the multiplicity and chirality of matter fields in a 4D
theory, including the genus of the matter curve, G-flux, and possibly
other considerations, our main intent here is to determine simply the
set of allowed representations, and we do not attempt to carry out a
precise calculation of multiplicity and chirality in specific 4D
models.  

In the next section we present a number of examples.  This has two
purposes.  First, in four-dimensional theories we describe explicit
geometries that realize  $SU(3)\times SU(2)$
non-Higgsable clusters and $SU(3)$ non-Higgsable
clusters with Higgsable $SU(2)$ factors,
in both cases with matter; this is an existence proof for
constructions of these types.  Second, by studying six-dimensional
theories where we engineer the $IV$-$IVm$, $IV$-$III$, and $IV$-$I_2$
intersections, we determine the set of nonzero charges that can arise in
the spectrum using anomaly cancellation.  Specifically, anomaly
cancellation  allows us to precisely determine the non-trivial
$SU(3)\times SU(2)$ matter representations localized at the collision
of the $SU(3)$ and $SU(2)$ seven-branes, including multiplicities in
the 6D theory, and we can match the set of  representations found in
that way with the junction analysis of this section. 

Of the three ways ($III$, $IVm$, $I_2$) that a $SU(2)$ can be realized
in conjunction with a non-Higgsable $SU(3)$, in this section we focus
on the $IV$-$III$ intersection.  The other possibilities are analyzed
in the following section.  A summary of the geometric matter realized
in these three configurations is given in \S\ref{sec:summary.spectra}.

\subsection{Non-perturbative Realizations \\ of $SU(3)$ and $SU(2)$}

We begin by reviewing the structure of $SU(2)$ and $SU(3)$ gauge
states that arise from a type $III$ and type $IV$ fiber, as described
by string junctions \cite{Schwarz-SJ,Gaberdiel-Zwiebach,
  DeWolfe-Zwiebach} since they differ significantly from the case of
$n$ D7-branes.  See \cite{GHS-II} for detailed deformations that give
rise to this data.  These junction descriptions are well understood
but set the stage for the more complicated junction descriptions of
intersecting seven-branes.

Before reviewing these $SU(3)$ and $SU(2)$ theories, we present a
brief basic review of string junctions and their connection to
deformations of algebraic varieties as developed\footnote{For computer
  codes which aid in performing these computations, see
  \url{http://www.jhhalverson.com/deformations}.}  in
\cite{GHS-I,GHS-II}. Consider an M-theory compactification to $(d-1)$
dimensions on an elliptically fibered Calabi-Yau manifold $X$ in
Weierstrass form
\begin{equation}
y^2 = x^3 + f\, x + g =: v_3(x).
\end{equation}
This theory becomes an F-theory compactification to $d$ dimensions in
its vanishing fiber limit $X\rightarrow X_{v}$. Suppose that $X$
exhibits a codimension one singular fiber at a locus $z=0$ where a
local coordinate $z$ on the base vanishes, giving a non-abelian gauge
factor $G$, so that $\Delta = z^n \Delta_r$ for some $n>2$. Then a
small disc transverse to $z=0$ defines an (open) elliptic surface in
$X$; this surface comes in a family that can be parameterized by local
complex coordinates on the locus $\{z = 0\}$.
If one performs a small deformation in a local patch by deforming $f$ and
$g$ such that $X\rightarrow X_{\epsilon}$  has discriminant $\Delta
= z^n \Delta_r + \epsilon \Delta_\epsilon$ with $\epsilon \in \bC$ and
$z$ not dividing $\Delta_\epsilon$, then the non-abelian gauge
symmetry along $z=0$ in $X$ is spontaneously broken. By taking the
limit $\epsilon\rightarrow 0$ the symmetry is restored.

Consider a generic elliptic surface in the family of elliptic surfaces
that cross the singular locus $z=0$ in $X$. In performing the deformation
$X\rightarrow X_{\epsilon}$ for small $\epsilon$, the $n$ marked
points at $z=0$ in the disc $D$ (which are seven-branes in the F-theory
limit) spread out into $n$ non-degenerate marked points $z_i$, each
with an $I_1$ singular fiber above it in the elliptic surface. Each
such fiber has an associated vanishing cycle, which is the $(p,q)$
label of the seven-brane at $z_i$ in F-theory. The fiber above $z=0$
in the deformed elliptic surface is smooth, and by following straight
line paths from $z=0$ to the $z_i$, the vanishing cycles can be read
off systematically. Specifically, for any point in $D\setminus \{ z_i\}$
the associated fiber is smooth
 and $v_3(x)$ has three non-degenerate roots
that might appear as the dots in
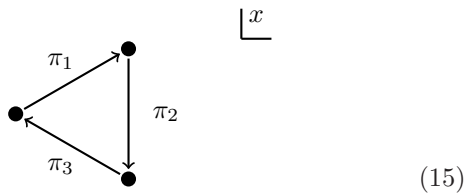
\begin{equation} 
\begin{tikzpicture}[scale=1]
    \fill[xshift=7cm,thick] (180:10mm) circle (1mm);
    \fill[xshift=7cm,thick] (180-120:10mm) circle (1mm);
    \fill[xshift=7cm,thick] (180+120:10mm) circle (1mm);
    \node at (9.2cm,1.3cm) {$x$};
    \draw[xshift=7cm,thick,->] (180:10mm)+(30:1.3mm) -- +(30:16mm);
    \draw[xshift=7cm,thick,->] (180-120:10mm)+(-90:1.3mm) -- +(-90:16mm);
    \draw[xshift=7cm,thick,->] (180+120:10mm)+(150:1.3mm) -- +(150:16mm);
    \node at (6.6cm,0.7cm) {$\pi_1$};
    \node at (8cm,0cm) {$\pi_2$};
    \node at (6.6cm,-0.7cm) {$\pi_3$};
    \draw[xshift=9cm,thick,yshift=1.0cm] (90:0mm) -- (90:4mm);
    \draw[xshift=9cm,thick,yshift=1.0cm] (0:0mm) -- (0:4mm);
 \end{tikzpicture}
\label{eqn:pixpiypiz}
\end{equation}
These points are where the double cover $y^2=v_3(x)$
degenerates. A one-cycle on the elliptic fiber can be defined by
following a path from one dot to another, going to the other sheet,
and then following a path back to the original dot; in the figure we
have defined three one-cycles $\pi_1$, $\pi_2$, $\pi_3$ subject to the
condition $\pi_1+\pi_2+\pi_3=0$. In following the mentioned straight
line path from $z=0$ to one of the $z_i$, two of these points will
collide as $z \rightarrow z_i$ and the corresponding one-cycle
vanishes in this limit. Picking a basis, this determines the $(p,q)$
label of the seven-brane at $z_i$, and in particular one can identify
an ordered set of vanishing cycles $Z=\{\pi_1,\cdots,\pi_n\}$.

As expected from string theory, objects can stretch between the branes
located at the $z_i$. Suppose that two branes at $z_1$ and $z_2$ have
the same vanishing one-cycle $\pi_1=\pi_2\equiv \pi$. Then in
following a path from $z_1$ to $z_2$, the vanishing one-cycle begins
as a point at $z_1$, grows, shrinks, and then finally collapses to a
point at $z_2$; this defines a two-sphere in the total space of the
elliptic surface with one leg on the fiber and one leg on the base,
and an M2-brane can be wrapped on this two-cycle in the M-theory
picture. In the F-theory limit this M2-brane becomes a fundamental
string (with an appropriate choice of $SL(2,\bZ)$ frame). If there
were three seven-branes, it is natural to represent the state from
$z_1$ to $z_2$ as $(1,-1,0)$, and the negative of this for the state
from $z_2$ to $z_1$.

More generally, though, if a collection of branes at the $z_i$ have
different vanishing cycles, it is possible to form two-spheres that
end on multiple branes, with an associated ``junction'' in the base
that generalizes the path from one $z_i$ to another. We will see what
this means in an example, momentarily; for now just note that we
represent such a junction by a tuple $(J_i)$ in $\bZ^n$ where $n$ is
the number of branes being analyzed in the deformation and $J_i$ is
the number of ``prongs'' coming off the brane at $z_i$. A junction $J$
may have a loose end that goes off to infinity with some asymptotic
charge $a(J)\equiv \sum J_i \pi_i$, which is just the $(p,q)$ charge
of the loose end (the one-cycle wrapped by an M2-brane in the M-theory
picture) emanating to infinity (perhaps ending on a far away
brane). Junctions with $a(J)$ zero end entirely within the branes at
the $z_i$, each defining a two-sphere on which M2-branes can be
wrapped in the M-theory picture; these become string junctions in the
F-theory limit. One definition of a string junction, then, is the
object obtained by taking the F-theory limit of an M2-brane on the
deformation two-cycle. Finally, the two-spheres associated to
junctions have non-trivial topological intersections in the elliptic
surface. In \cite{GHS-I} we wrote down a formula for these
intersections, which determines a so-called $I$-matrix that
conveniently computes the intersections.

From this data there are various computations that can be
performed. For example, if the deformed branes are associated to a
Kodaira singularity with Lie algebra $G$ (which may or may not be the
gauge group, depending on details), then the junctions with $a(J)$
zero and self-intersection $-2$ are determine the root lattice of $G$.

Having reviewed the basics of junctions, let us turn to examples, which
may provide further clarification.

\vspace{.7cm}
\noindent \emph{$SU(2)$ from a type $III$ fiber and Mercedes W-bosons.}  

A seven-brane with gauge symmetry $SU(2)$ can arise from a type $III$
fiber; if it is along $z=0$ and specified by a Weierstrass model, then
$({\rm ord}(f), {\rm ord}(g), {\rm ord}(\Delta))=(1,2,3)$.  Note that
since ${\rm ord}(\Delta) = 3$, this $SU(2)$ gauge symmetry arises when
$\emph{three}$ seven-branes collide, as opposed to the common ($I_2$)
case in weakly coupled type IIB that arises when two D7-branes
collide.  In \cite{GHS-II} it was shown that the type $III$
configuration can arise from three seven-branes with $(p,q)$ labels
\begin{equation}
Z_{III} = \{ \pi_2, \pi_1, \pi_3\}
\end{equation}
which has associated $I$-matrix
\begin{equation}
  I = (\cdot, \cdot) = \begin{pmatrix} -1 & 1/2 & -1/2 \\ 1/2 & -1 & 1/2 \\ -1/2 & 1/2 & -1\end{pmatrix}.
\end{equation}
The topological self intersections of junctions can be determined from
this matrix.  Defining the asymptotic charge $a(J) = \sum J_i \pi_i\in
H_1(T^2,\bZ)$, the set $R = \{J \in \bZ^3 \,\,\, | \,\,\, (J,J) =
-2\,\, \text{and} \,\,a(J)=0\}$ is computed to be
$R=\{(1,1,1),(-1,-1,-1)\}$.  These are three-pronged string junctions
\begin{equation}
  \begin{tikzpicture}
  \fill[xshift=-20mm] (90:8mm) circle (1mm);
  \fill[xshift=-20mm] (210:8mm) circle (1mm);
  \fill[xshift=-20mm] (330:8mm) circle (1mm);
  \draw[xshift=-20mm,thick] (0:0mm) -- (90:7mm);
  \draw[xshift=-20mm,thick] (90:3.5mm) -- (75:3mm);
  \draw[xshift=-20mm,thick] (90:3.5mm) -- (105:3mm);
  \draw[xshift=-20mm,thick] (0:0mm) -- (210:7mm);
  \draw[xshift=-20mm,thick] (210:3.5mm) -- (225:3mm);
  \draw[xshift=-20mm,thick] (210:3.5mm) -- (195:3mm);
  \draw[xshift=-20mm,thick] (0:0mm) -- (330:7mm); 
  \draw[xshift=-20mm,thick] (330:3.5mm) -- (345:3mm);
  \draw[xshift=-20mm,thick] (330:3.5mm) -- (315:3mm);

  \fill[xshift=20mm] (90:8mm) circle (1mm);
  \fill[xshift=20mm] (210:8mm) circle (1mm);
  \fill[xshift=20mm] (330:8mm) circle (1mm);
  \draw[xshift=20mm,thick] (90:0mm) -- (90:7mm);
  \draw[xshift=20mm,thick] (90:3.5mm) -- (78:4.1mm);
  \draw[xshift=20mm,thick] (90:3.5mm) -- (102:4.1mm);
  \draw[xshift=20mm,thick] (0:0mm) -- (210:7mm);
  \draw[xshift=20mm,thick] (210:3.5mm) -- (222:4.1mm);
  \draw[xshift=20mm,thick] (210:3.5mm) -- (198:4.1mm);
  \draw[xshift=20mm,thick] (0:0mm) -- (330:7mm); 
  \draw[xshift=20mm,thick] (330:3.5mm) -- (342:4.1mm);
  \draw[xshift=20mm,thick] (330:3.5mm) -- (318:4.1mm);
  \end{tikzpicture}
\end{equation}
that are the $W_+$ and $W_-$ bosons of $SU(2)$, where the charge is
with respect to the Cartan $U(1)$ of $SU(2)$; in the M-theory picture
one dimension lower these are M2-branes wrapped on the corresponding
two-cycles.  In \cite{GHS-II} it was shown how to build up higher spin
representations of $SU(2)$ using this junction data.

\vspace{.3cm}
\noindent \emph{$SU(3)$ from a type $IV$ fiber.}  

A seven-brane with gauge symmetry $SU(3)$ can arise from a type $IV$
fiber; if it is along $z=0$ and specified by a Weierstrass model, then
$({\rm ord}(f),{\rm ord}(g),{\rm ord}(\Delta)) = (2,2,4)$.  Again we see that there are
more seven-branes than expected from type IIB expectations; in the
perturbative IIB case, $SU(3)$ arises from a stack of three coincident D7-branes,
{\it i.e.}  from a type $I_3$ fiber, whereas for a type $IV$ fiber $SU(3)$ is
realized by \emph{four} seven-branes.  In \cite{GHS-II} it was
shown that this type $IV$ configuration can arise from four seven-branes
with $(p,q)$ 
labels
\begin{equation}
  Z_{IV} = \{\pi_1,\pi_3,\pi_1,\pi_3\}.
\end{equation}
and associated $I$-matrix
\begin{equation}
I = (\cdot, \cdot) = \begin{pmatrix}
-1 & 1/2 & 0 & 1/2 \\
1/2 & -1 & -1/2 & 0 \\
0 & -1/2 & -1 & 1/2 \\
1/2 & 0 & 1/2 & -1
\end{pmatrix}.
\end{equation}
Computing (as we did for the type $III$ case) the set of junctions
with self-intersection $-2$ and asymptotic charge zero, we obtain
\begin{align}
\{(-1, -1, 1, 1), (-1, 0, 1, 0), (0, -1, 0, 1), \nonumber\\(0, 1, 0, -1), (1, 0, -1, 0), (1, 1, -1, -1)\},
\end{align}
a set of six string junctions that fills out (together with the
Cartan elements, which are two copies of $(0,0,0,0)$) an adjoint of
$SU(3)$.   See \cite{GHS-II} for a detailed description of simple roots
and how to build arbitrary $SU(3)$ representations using this junction
data.

\subsection{$SU(3)\times SU(2)$ from a $IV$-$III$ collision}

We would like to study the geometry and physics at the $IV$-$III$
intersection; we are interested in particular in the matter spectrum that
occurs there.  
In a simple $I_3 \times I_2$ collision, the matter fields consist
simply of bifundamental matter that transforms in the $({\bf 3}, {\bf 2})$ of
$SU(3) \times SU(2)$, associated with open strings that connect the
three D7-branes in the $I_3$ to the two D7-branes in the $I_2$.  We
expect to find a similar $({\bf 3}, {\bf 2})$ matter representation in other
constructions of $SU(3) \times SU(2)$ on intersecting divisors.  If
these gauge groups are those of the standard model, then the natural
interpretation of the $({\bf 3}, {\bf 2})$ matter is as the left-handed quark
doublets.  

The surprising feature of the $IV$-$III$ intersection is that we also
get fields that transform in the same representations as the other
matter fields of the standard model. Note that the purely geometric analysis always
describes non-chiral (``${\cal N} = 2$'') matter, so that for every
representation such as $({\bf 3}, {\bf 2})$ there is a matter field of opposite
chirality in the conjugate representation $(\bar{{\bf 3}}, \bar{{\bf
    2}})$ (though $\bar{{\bf 2}} = {\bf 2}$).  In much of the ensuing
discussion we do not explicitly mention these conjugate
representations, assuming throughout that all matter arising from
purely geometric analyses is non-chiral.  The reduction through G-flux
to chiral matter is discussed in \S\ref{sec:issues}.

We first review the structure of matter realized at a perturbative
$SU(3)\times SU(2)$ intersection realized by intersecting D7-branes.
We then present a string junction analysis in neighborhood of the
collision; this is not a full deformation analysis of string junctions
but gives intuition for why the states localized at the collision are
richer than in the perturbative case.  We determine the precise
charged $SU(3)\times SU(2)$ spectrum at a $IV$-$III$ collision in the
next section via a six-dimensional anomaly analysis.  Combining these
analyses gives us a complete list of the kinds of representations that
can appear at this intersection.  As mentioned above, a complete
treatment of multiplicities and chirality would require a more
thorough analysis that we leave for later work.  Note, however, that
the analysis of this section applies to the matter localized at any
$IV$-$III$ collision, regardless of Higgsability or non-Higgsability
and independent of dimension.

\sskip
\noindent \emph{``Quark Doublets'' From an $I_3$-$I_2$ Collision.}
\vsskip

We begin with the open string/string junctions description of a well
known case that will be in sharp contrast with the $IV$-$III$
geometry.  Consider a stack of three D7-branes along $z=0$ that has
transverse intersection with a stack of two D7-branes along $t=0$.
This is a type IIB configuration with $SU(3)\times SU(2)$ gauge
symmetry; the strings localized at $z=t=0$ transform in the
bifundamental representation $({\bf 3}, {\bf 2})$, which can become quark doublets
$({\bf 3}, {\bf 2})_{1/6}$ after turning on worldvolume fluxes and engineering an
embedding of the weak hypercharge.  This type IIB configuration is
equivalent to the an F-theory geometry that realizes an $I_3$ fiber
along $z=0$ and an $I_2$ fiber along $t=0$, in which case
\begin{equation}
\Delta = z^3t^2\, \tilde \Delta
\end{equation}
where $\tilde \Delta$ is a residual piece of the discriminant that is
trivially computed in examples with a toric base.

\begin{figure}[t]
\begin{tikzpicture}
  \node [xshift=-20mm,yshift=15mm] at (0,0) {$I_3$-$I_2$ Collision};
  \draw [xshift=-20mm,yshift=13mm] (0:-12.5mm)--(0:12.5mm);
  \draw [xshift=-20mm,yshift=13mm] (0:-12.5mm)+(0,3mm)--(0:-12.5mm);
  \draw [xshift=-20mm,yshift=13mm] (0:12.5mm)+(0,3mm)--(0:12.5mm);
  \node [xshift=22mm,yshift=15mm] at (0,0) {$IV$-$III$ Collision};
  \draw [xshift=22mm,yshift=13mm] (0:-14mm)--(0:14mm);
  \draw [xshift=22mm,yshift=13mm] (0:-14mm)+(0,3mm)--(0:-14mm);
  \draw [xshift=22mm,yshift=13mm] (0:14mm)+(0,3mm)--(0:14mm);
  \draw [xshift=-20mm,thick] (0:-10mm)--(0:10mm);
  \draw [xshift=-20mm,thick] (90:-10mm)--(90:10mm);
  \draw [xshift=-20mm,thick, dotted] (110:8mm) -- (-20:8mm);
  \node [xshift=-20mm] at (-90:12mm) {$2$};
  \node [xshift=-20mm] at (0:13mm) {$3$};
  \draw [xshift=20mm,thick] (0:-10mm)--(0:10mm);
  \draw [xshift=20mm,thick] (90:-10mm)--(90:10mm);
  \draw [xshift=20mm,thick, dotted] (110:8mm) -- (-20:8mm);
  \node [xshift=20mm] at (-90:12mm) {$3$};
  \node [xshift=20mm] at (0:13mm) {$4$};
  \node [xshift=20mm] at (130:13mm) {$\tilde \Delta$};
  \node [xshift=20mm] at (50:13mm) {$1$};
  \draw [xshift=20mm,thick] (0,0) parabola (8mm,8mm);
  \draw [xshift=20mm,thick] (0,0) parabola (-8mm,8mm);
\end{tikzpicture}
\caption{Displayed are the local geometries of an $I_3$-$I_2$ collision
  and a $IV$-$III$ collision, respectively.  In both cases the $SU(2)$
  gauge theory is on the vertical line, the $SU(3)$ gauge theory is on
  the horizontal line, and quark doublets are localized at the
  intersection of the solid lines; the integers denote the number of
  branes in each stack.  Note the additional brane $\tilde \Delta$
  participating in the intersection in the $IV$-$III$ case.}
\label{figure:intersections}
\end{figure}
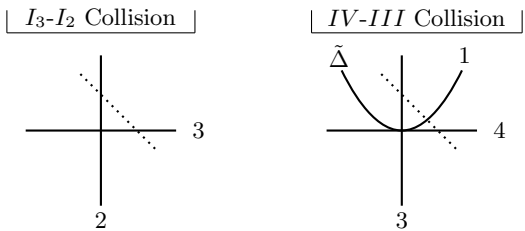

The matter content of the theory can be determined by resolving the
geometry or by studying a deformation of the geometry near the
intersection of the branes.   A first step in the case of a full
deformation analysis is to study properties of the deformed geometry
in an elliptic surface near the collision; this is the analysis we
will perform.   To do this, we choose an appropriate cross-section of
the branes, as depicted for example by the dotted line on the left
hand side of Figure \ref{figure:intersections}.  A deformation of the
geometry can yield the picture
\begin{equation}
  \begin{tikzpicture}
    \fill[xshift=5mm,color=black] (0:6mm) circle (1mm);
    \fill[xshift=5mm,color=black] (90:4mm) circle (1mm);
    \fill[xshift=5mm,color=black] (270:4mm) circle (1mm);
    \fill[xshift=-30mm,color=black] (90:4mm) circle (1mm);
    \fill[xshift=-30mm,color=black] (270:4mm) circle (1mm);
    \node at (8mm,-10mm) {$SU(3)$ Simple Roots};
    \node at (8mm,-14mm) {from Deformed $I_3$};
    \node at (-30mm,-10mm) {$SU(2)$ Simple Roots};
    \node at (-30mm,-14mm) {from Deformed $I_2$};
    \node at (-33mm,0mm) {$\alpha_1$};
    \node at (-12mm,7mm) {$\alpha_2$};
    \node at (10mm,5mm) {$\alpha_3$};
    \node at (10mm,-5mm) {$\alpha_4$};
    \draw [xshift=-30mm,thick,<-] (90:6mm)+(270:3mm) -- (270:4mm);
    \draw [xshift=5mm,thick,<-] (0:6mm)+(150:1.1mm) -- (90:4mm);
    \draw [xshift=5mm,thick,<-] (270:4mm)+(45:1.1mm) -- (0:6mm);
    \draw [xshift=-30mm,thick,dashed,<-] (6.8:34mm) -- (90:4mm);
 \end{tikzpicture}
\end{equation}
where the page describes the complex plane along the mentioned dotted
line, and we note that the branes of the $SU(2)$ and $SU(3)$ gauge
theories are now split.  

Specifically, both gauge factors have been spontaneously broken, which
allows us to study the now-massive states in the deformed geometry.
Above, the arrows are string states that can be represented as vectors
in $\bZ^5$, specifically
\begin{equation}\alpha_i = e_i - e_{i+1}\end{equation}
for an orthonormal basis $e_i\in \bZ^5$.  The set of states
$\{\alpha_1,\alpha_2,\alpha_3,\alpha_4\}$ are simple roots of $SU(5)$
and thus generate an entire adjoint of $SU(5)$.  

In the undeformed geometry all of these states become massless at
$z=t=0$ since the branes collide; only those states in the adjoint of
$SU(5)$ that are in the adjoint of $SU(2)$ are massless along the
entire $t=0$ locus, and similarly for $SU(3)$ adjoint states along
$z=0$.  Thus, the states that become massless at only $z=t=0$ are the
ones that have one end on each of the different stacks of branes;
these are roots of $SU(5)$ that have a contribution from $\alpha_2$,
the simple root represented by the dotted line above.  These string
states are
\begin{center}
  $\alpha_2, \,\, \alpha_2 + \alpha_3, \,\, \alpha_2 + \alpha_3 + \alpha_4$ \\
  $\alpha_1 + \alpha_2, \,\, \alpha_1 + \alpha_2 + \alpha_3, \,\,
  \alpha_1 + \alpha_2 + \alpha_3 + \alpha_4$
\end{center}
together with their negatives, and they fill out the (reducible)
representation
$(3,\ov 2) \oplus (\ov 3,2)$ of $SU(3) \times SU(2)$, where the
difference between $\ov 2$ and $2$ is the sign of the arrow into the
$SU(2)$ stack.  This result matches the branching rule of an $SU(5)$
adjoint
\begin{equation}
  24 \longrightarrow (8,1)  \oplus  (1,3)  \oplus (3,\ov 2) \oplus (\ov 3,2) \oplus (1,1) 
\end{equation}
into representations of $SU(3)\times SU(2)$, in accord with the
Katz-Vafa procedure.  In summary, a $(3,\ov 2)
\oplus (\ov 3, 2)$ of $SU(3)\times SU(2)$ is localized at $z=t=0$,
which can in principle
become a chiral supermultiplet of left-handed quark doublets 
after turning on flux and engineering an embedding of weak hypercharge.

There are a few particular features we would like to note about this
well-known example.
\begin{itemize}
\item The only branes at the $SU(3)\times SU(2)$ intersection are the
  $SU(3)$ and $SU(2)$ brane stacks themselves.
\item The states that become massless at the intersection are
  charged under both $SU(3)$ and $SU(2)$.
\item Obtaining this geometry requires specializing in moduli space.
  Obtaining additional states --- such as the other standard model
  fields --- may require an additional specialization in moduli
  space.
\end{itemize}
We will see the first two of these statements are not true of the $IV$-$III$
geometry.  Furthermore, the third is not true when the $IV$-$III$ collision
is non-Higgsable.

\sskip
\noindent \emph{Standard Model Matter Representations \\ From a $IV$-$III$ Collision.}
\vsskip

We now turn to study the $IV$-$III$ geometry.  As mentioned above, the
main point of this section is that if one sets out to obtain quark
doublets via colliding a type $III$ fiber with a type $IV$ fiber, one
obtains not only these states, but also $SU(3)\times SU(2)$
representations that may realize
the other standard model
matter fields.  Moreover, these states are all localized on the
\emph{same matter curve} in the threefold base.

Consider an F-theory model that realizes $SU(3)\times SU(2)$ gauge
symmetry  with a
type $IV$ and type $III$ fiber, respectively, rather than the usual
$I_3$ and $I_2$ of the type $IIB$ string.   A Weierstrass model that
realizes this possibility must be specified by the data
\begin{equation}
\label{eq:IV-IIIWeierstrass}
f = z t^2\, \tilde f \qquad g = z^2 t^2 \, \tilde g 
\end{equation}
where $z$, $t$, $\tilde f$ and $\tilde g$ are (local) sections of
$\cO(Z)$, $\cO(T)$, $\cO(-4K_B-Z-2T)$, and $\cO(-6K_B-2(Z+T))$,
respectively.  The discriminant takes the form
\begin{align}
  \Delta = z^3 t^4 \left(4 t^2\, \tilde f^3 +27z\, \tilde g^2 \right)
  \equiv z^3 t^4\, \tilde \Delta.
\end{align}
where the residual discriminant $\tilde \Delta$ is a section of
$\cO(-12K_B -3Z-4T)$.   Assuming there is no outer monodromy on the
type $IV$ fiber, which holds if $g = t^2 g_2 + \cO(t^3)$ has $g_2$ a
perfect square, then the seven-branes along $t=0$ and $z=0$ have gauge
symmetry $SU(3)$ and $SU(2)$, respectively.  We assume that the
codimension two locus $z=t=0$ exists in the geometry, since this is
is necessary to have the collision, and therefore quark doublets; in
section \ref{sec:examples} we present explicit examples realizing this possibility.

As in the case of the $I_3$-$I_2$ intersection, the seven-branes
carrying $SU(3)$ and $SU(2)$ gauge theories intersect at $z=t=0$,
where one expects the localization of massless charged states.  There
are two important differences, however.  First, the $SU(3)$ and
$SU(2)$ seven-branes are stacks of four and three $(p,q)$
seven-branes, rather than stacks of three and two D7-branes; thus, the
$SU(3)$ and $SU(2)$ stacks contribute more branes to the $z=t=0$ locus
in a $IV$-$III$ collision than in an $I_3$-$I_2$ collision.  Second,
in the case of a $IV$-$III$ collision it is easy to see from the $z,t$
dependence of $\tilde \Delta$ that $\tilde \Delta=0$ \emph{always}
intersects $z=t=0$, unlike in an $I_3$-$I_2$ collision; {\it i.e.},  there is
always an additional brane, henceforth called the \emph{extra brane},
and we will see that charged states may end on it.  This corresponds
to the observation mentioned earlier that the intersection of loci
with orders of vanishing of $f, g, \Delta$ of $(1, 2, 3)$ and $(2, 2,
4)$ must have vanishing orders $(3, 4, 8)$, not $(3, 4, 7)$.  
This enhancement of the order of vanishing of $\Delta$
comes from an additional $I_1$ codimension one singularity that must
also meet the $IV$-$III$ intersection point.
Thus, this intersection point might more properly be referred to as a
$IV$-$III$(-$I_1$) collision, with the parentheses indicating that the
third brane is automatically produced when the first two singularities collide.
The
structure of the two intersecting geometries near the codimension two
intersection is given in Figure \ref{figure:intersections}.

The vanishing degrees $(3, 4, 8)$ of $f, g, \Delta$ at the
intersection point corresponds to a Kodaira type $IV^{*}$ singularity,
associated with an $E_6$ Dynkin diagram.  We thus expect that the
matter fields at the singularity should be those associated with a
decomposition of the adjoint of $E_6$ when $SU(3) \times SU(2)$ is
embedded as a subgroup.  The deformation analysis using junctions
provides a simple way of seeing explicitly how this decomposition works.

\sskip Let us therefore
now deform the geometry in order to study the
$SU(3)\times SU(2)$ representations of the states localized at the
$IV$-$III$(-$I_1$) collision.  Consider a compact elliptically fibered
Calabi-Yau variety in the special Weierstrass form
(\ref{eq:IV-IIIWeierstrass})  in a patch that gives
a local model near the $IV$-$III$ intersection.  Consider the deformation
of the local model defined by
\begin{equation}
(f,g) \longrightarrow (f, g+ \epsilon),
\end{equation}
where $\epsilon$
may be the restriction of a deformation of the
global model, depending on the example; alternatively it may simply be
considered as a deformation of the local model.  This deforms the
discriminant to
\begin{equation}
\Delta = 4  t^6 \tilde f^3 z^3+27 \left( t^2 z^2 \tilde g\, +\epsilon \right)^2.
\end{equation}
and we see that the three and four seven-branes along $z=0$ and $t=0$,
respectively, have split up; the $z=0$ and $t=0$ singular loci have been
smoothed.

Given this deformation, we would like to study an elliptic surface
nearby the $IV$-$III$(-$I_1$) collision by taking a transverse slice, as
displayed in Figure \ref{figure:intersections}.  Taking $\tilde f = \tilde g=1$ and defining the local
coordinate
\begin{equation}
p_\perp = t - \epsilon_z\, z-t_{0},
\end{equation}
where $\epsilon_z\in \bC$ is a rotation factor and $t_0$ is an offset,
we can define a family of transverse slices of the local geometry by
$p_\perp = 0$ for various values of the parameters.  A generic pair
$(t_0,\epsilon_z)$ should suffice; we take $(5,2e^{i \pi / 5})$ and
$\epsilon=10$, in which case the deformed seven-branes intersect the
transverse plane as
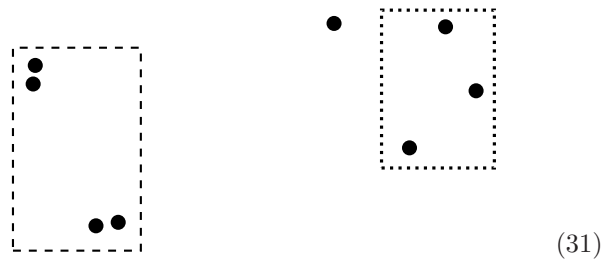
\begin{equation}
\begin{tikzpicture}
  \fill[thick,xshift=45mm] (-13.5791/2.5, -0.456807/2.5) circle (1mm); 
  \fill[thick,xshift=45mm] (-13.5084/2.5, 0.1615734/2.5) circle (1mm); 
  \fill[thick,xshift=45mm] (-11.4908/2.5, -5.17381/2.5) circle (1mm); 
  \fill[thick,xshift=45mm] (-10.7521/2.5, -5.05719/2.5) circle (1mm); 
  \fill[thick,xshift=45mm] (-3.57851/2.5, 1.55766/2.5) circle (1mm); 
  \fill[thick,xshift=45mm] (-1.07091/2.5, -2.58071/2.5) circle (1mm); 
  \fill[thick,xshift=45mm] (0.126672/2.5, 1.44483/2.5) circle (1mm); 
  \fill[thick,xshift=45mm] (1.14206/2.5, -0.682583/2.5) circle (1mm);
  \draw[thick,dashed] (-12mm,3mm)--(-12mm,-24mm)--(5mm,-24mm)--(5mm,3mm)--cycle;
  \draw[very thick,dotted] (37mm,8mm)--(37mm,-13mm)--(52mm,-13mm)--(52mm,8mm)--cycle;
\end{tikzpicture}
\end{equation}
where the dashed box contains the split $SU(3)$ branes of the type
$IV$ fiber, the dotted box contains the split $SU(2)$ branes of the
type $III$ fiber, and the seven-brane on its own is the deformed extra
brane.  Using the techniques of \cite{GHS-I,GHS-II} to read off the
vanishing one-cycles in the elliptic fiber above these marked points,
beginning with the extra brane and moving clockwise, we find
\begin{equation}
Z_{IV-III} = \{\pi_2,\pi_1,\pi_3,\pi_2,\pi_1,\pi_3,\pi_1,\pi_3 \} 
\end{equation}
which has associated I-matrix
\begin{equation}
I = (\cdot, \cdot) = 
\begin{pmatrix}
-1& \frac{1}{2}& -\frac{1}{2}& 0& \frac{1}{2}& -\frac{1}{2}& \frac{1}{2}& -\frac{1}{2} \\ 
\frac{1}{2}& -1& \frac{1}{2}& -\frac{1}{2}& 0& \frac{1}{2}& 0& \frac{1}{2} \\
-\frac{1}{2}& \frac{1}{2}& -1& \frac{1}{2}& -\frac{1}{2}& 0& -\frac{1}{2}& 0 \\ 
0& -\frac{1}{2}& \frac{1}{2}& -1& \frac{1}{2}& -\frac{1}{2}& \frac{1}{2}& -\frac{1}{2} \\ 
\frac{1}{2}& 0& -\frac{1}{2}& \frac{1}{2}& -1& \frac{1}{2}& 0& \frac{1}{2} \\
-\frac{1}{2}& \frac{1}{2}& 0& -\frac{1}{2}& \frac{1}{2}& -1& -\frac{1}{2}& 0 \\ 
\frac{1}{2}& 0& -\frac{1}{2}& \frac{1}{2}& 0& -\frac{1}{2}& -1& \frac{1}{2} \\
-\frac{1}{2}& \frac{1}{2}& 0& -\frac{1}{2}& \frac{1}{2}& 0& \frac{1}{2}& -1 \\
\end{pmatrix}.
\end{equation}
This is all of the data necessary to perform a first analysis of the string junctions
near the collision.

Let us study the roots realized locally in the geometry by computing
$\{J \in \bZ^8 \,\, | \,\, (J,J) = -2 \,\,\, \text{and} \,\,\, a(J) =
0\}$.  We find that it is composed of $72$ junctions that fill out
(together with the six Cartan elements $(0,0,0,0,0,0,0,0)$) an adjoint
of $E_6$.  A set of simple roots is given by
\begin{align}
\alpha_1 &= (0, 0, 0, 0, 1, 0, -1, 0) \nonumber \\
\alpha_2 &= (0, 0, 0, 0, 0, 1, 0, -1) \nonumber \\
\alpha_3 &= (0, 0, 0, -1, -1, -1, 0, 0) \nonumber \\
\alpha_4 &= (0, 1, 1, 1, 0, 0, 0, 0)\nonumber \\
\alpha_5 &= (0, -1, 0, 1, 1, 0, 1, 1)\nonumber \\
\alpha_6 &= (-1, -1, -1, 0, 0, 0, 0, 0)
\end{align}
Note that $\alpha_1$ and $\alpha_2$ are string junctions that end only
on the last four seven-branes; these are the simple roots of the
$SU(3)$ gauge theory.    Similarly, $\alpha_4$ ends only on the second,
third, and fourth, seven-brane; these are the simple roots of the
$SU(2)$ gauge theory.  The only simple root that ends on the extra
brane is $\alpha_6$.

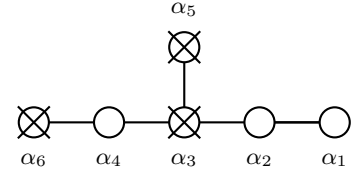
\begin{figure}
  \centering
  \begin{tikzpicture}
    \draw [thick] (0mm,0mm) circle (2mm);
    \draw [thick] (-10mm,0mm) circle (2mm);
    \draw [thick] (-20mm,0mm) circle (2mm);
    \draw [thick] (-20mm,10mm) circle (2mm);
    \draw [thick] (-30mm,0mm) circle (2mm);
    \draw [thick] (-40mm,0mm) circle (2mm);
    \draw [thick] (-2mm,0mm) -- (-8mm,0mm);
    \draw [thick] (-2mm,0mm) -- (-8mm,0mm);
    \draw [thick] (-12mm,0mm) -- (-18mm,0mm);
    \draw [thick] (-22mm,0mm) -- (-28mm,0mm);
    \draw [thick] (-20mm,2mm) -- (-20mm,8mm);
    \draw [thick] (-32mm,0mm) -- (-38mm,0mm);
    \node at (0mm,-5mm) {$\alpha_1$};
    \node at (-10mm,-5mm) {$\alpha_2$};
    \node at (-20mm,-5mm) {$\alpha_3$};
    \node at (-30mm,-5mm) {$\alpha_4$};
    \node at (-40mm,-5mm) {$\alpha_6$};
    \node at (-20mm,15mm) {$\alpha_5$};
    \draw [xshift=-40mm,thick] (45:3mm) -- (180+45:3mm);
    \draw [xshift=-40mm,thick] (-45:3mm) -- (180-45:3mm);
    \draw [xshift=-20mm,thick] (45:3mm) -- (180+45:3mm);
    \draw [xshift=-20mm,thick] (-45:3mm) -- (180-45:3mm);
    \draw [xshift=-20mm,yshift=10mm,thick] (45:3mm) -- (180+45:3mm);
    \draw [xshift=-20mm,yshift=10mm,thick] (-45:3mm) -- (180-45:3mm);
  \end{tikzpicture}
  \caption{In an elliptic surface near a $IV$-$III$(-$I_1$) collision the geometry exhibits an $E_6$
    singularity, where the the $SU(3)$ and $SU(2)$ Dynkin diagram
    embed as displayed above.  The crossed nodes are simple roots of
    $E_6$ whose associated states only become massless in codimension
    two.  The left-most node corresponds to the $E_6$ simple root
    junction charged under the extra brane.}
\label{fig:E6DynkinWithBreaking}
\end{figure}

This is an entire adjoint of $E_6$ in an elliptic surface near the
$IV$-$III$(-$I_1$) collision, and accordingly those junctions in the adjoint
that aren't roots of $SU(3)$ or $SU(2)$ may give rise to matter
multiplets.  Since determining the $SU(3)\times SU(2)$ content of these
states ultimately amounts to branching the adjoint of $E_6$, we must
determine how $SU(3)$ and $SU(2)$ embed into $E_6$.  To do this, we
compute
\begin{equation}
-C_{ij}=(\alpha_i,\alpha_j) = \begin{pmatrix}
-2& 1& 0& 0& 0& 0 \\
1& -2& 1& 0& 0& 0 \\
0& 1& -2& 1& 1& 0 \\
0& 0& 1& -2& 0& 1 \\
0& 0& 1& 0& -2& 0 \\
0& 0& 0& 1& 0& -2
\end{pmatrix}
\end{equation}
and find that it is the negative Cartan matrix.
From this we construct the associated Dynkin diagram in Figure
\ref{fig:E6DynkinWithBreaking}, crossing out any node not associated
with the $SU(3)\times SU(2)$ gauge symmetry.   

The complete branching to $SU(3)\times SU(2)$ of this adjoint of $E_6$
is straightforward to compute.  A detailed list of the states in
resulting decomposition is given in Table \ref{table:IV-III Reps}.
Besides the junctions in the adjoints of $SU(3)$ and $SU(2)$, there
are junctions that transform in representations $({\bf 3}, {\bf 2}),
({\bf 3}, {\bf 1}),({\bf 1}, {\bf 2}),$ and $({\bf 1}, {\bf 1})$ (as
well as, the usual conjugates of the first two representations where
${{\bf 3}}$ is replaced by $\overline{{\bf 3}}$).  The junctions that
end on the extra brane thus fill out precisely the set of $SU(3)\times
SU(2)$ representations that are needed for the standard model matter
fields, including the singlet representation needed for the
right-handed lepton sector.

This is not a full analysis, and is not expected to reproduce
the multiplicities of the matter fields arising at the intersection;
for a given intersection of this form the deformation analysis could
be extended to include a continuous family of surfaces in the vicinity
of the singularity, and monodromy around one or the other of the
branes can in general lead to an identification between sets of states
that would reduce the multiplicity \cite{GHS-I}.  Indeed, we expect
such a reduction in this case, since the charged spectrum with
multiplicities computed in the 6D case in the following section shows
only one field in each of the distinct representations under $SU(3)
\times SU(2)$.  One important lesson that we take from this
deformation analysis, however, is that in addition to the charged
fields there are also fields that transform in the singlet
representation at the intersection point.

There is a simple way to understand why junctions with the
$SU(3)\times SU(2)$ quantum numbers of the standard model fermions
 have appeared in this analysis.
Notice that while we have a full adjoint of $E_6$ in the elliptic
surface near the $IV$-$III$(-$I_1$) collision, we have $SU(3)\times SU(2)$
gauge symmetry, and one could study how the $E_6$ states branch under
the intermediate $SU(5)$ or $SO(10)$ whose corresponding simple roots
are
\begin{align}
  SU(5) \text{ Simple Roots:}  \,\,\,\,\,\,\,\,&\{\alpha_1,\alpha_2,\alpha_3,\alpha_4\}\nonumber \\
  SO(10) \text{ Simple Roots:} \,\,\,\,\,\,\,\,&\{
  \alpha_1,\alpha_2,\alpha_3,\alpha_4,\alpha_5\}.
\end{align}
Note that $\alpha_6$ is the simple root lost in the $E_6\rightarrow
SO(10)$ branching, and that this is the only simple root junction
ending on the extra brane.  Accordingly, the states in the $E_6$
adjoint that end on the extra brane are those that are not roots of
$SO(10)$, which  therefore obey the
branching rule
\begin{equation}
72 \rightarrow 45 + 16 + 16' + 1
\end{equation}
and fit into a $16+16'$ of $SO(10)$.  Indeed this can be checked, and
that the further branching of those junctions to $SU(3)\times SU(2)$
respect the decomposition of the $16$ spinor of $SO(10)$ into standard
model representations.  Thus, the representation theory of $SO(10)$ plays
an important role in this geometry, even though the geometry does not
exhibit an $SO(10)$ GUT.

\sskip

We summarize by restating the result of this analysis: the 
$SU(3)\times SU(2)$ matter representations at the $IV$-$III$(-$I_1$) collision
include fields in the representations
\begin{equation}
(\fund_3,\fund_2), \; \fund_3, \; \fund_2, \; {\bf 1}.
\label{eq:junction-representations}
\end{equation}
While the slice-based junction analysis we performed in this section
does not nail down the multiplicities of these representations, the 6D
anomaly analysis in the following section shows that each of the
non-trivial representations appears with multiplicity one at the
intersection point.  The anomaly analysis, however, is not sensitive
to the singlet representations identified by the junction analysis.
The representations in (\ref{eq:junction-representations}) are
precisely the $SU(3) \times SU(2)$ representations of the various
matter fields in the standard model; thus, as we discussed further
below, in models with appropriately engineered $G$-flux and weak
hypercharge, it is possible that geometries of this type might realize
all of the standard model fermions along the same matter curve.

\section{Examples}
\label{sec:examples}

In this section we provide some specific examples of F-theory
compactifications giving type $IV$ and type $III$ singularities to
show how some of the ideas of this paper are realized in concrete
situations.   We begin in \S\ref{sec:examples.6} with some simple
examples in six-dimensional theories.   In \S\ref{sec:examples.IV} we
give two examples of 4D F-theory models with non-Higgsable $SU(3)$
gauge groups from Kodaira type $IV$ singularities, and two further
examples that have non-Higgsable $SU(3)$ gauge groups and Higgsable
$SU(2)$ gauge group factors.   In \S\ref{sec:examples.IV-III}, we give
an example of a 4D F-theory model that has a non-Higgsable $SU(3)
\times SU(2)$ gauge group from intersecting type $IV$ and type $III$
singularities and an $SU(3) \times SU(2)$
example of type $IV$-$IVm$.  These examples illustrate the structures involved in
the different geometric realizations of the gauge group factors, and
the different approaches that can be used to identify non-Higgsable
structure in given compactification geometries.

\subsection{Warm-up: some 6D examples} 
\label{sec:examples.6} 

We begin with a few simple examples of F-theory compactifications to
six dimensions with type $III$ and type $IV$ singularities.  In six
dimensions, as discussed earlier, the matter spectrum of an F-theory compactification is
highly constrained, and in many situations uniquely determined, by the
cancellation of gauge, gravitational, and mixed gauge-gravitational
anomalies, and there are no complications such as
G-flux that modify the spectrum beyond that determined by the geometry
of the F-theory compactification.

In the simplest of these examples, where there are no non-Higgsable
gauge group factors, we take the F-theory base manifold to be $B_2 =
\P^2$, and tune all codimension one singularities on linear $\P^1$'s
(complex lines) associated with the vanishing of coordinates $z, \yy$
(in a homogeneous coordinate system on $\P^2$ $[s: \yy: z]$).  Anomaly
cancellation uniquely determines the matter spectrum of an $SU(2)$ or
$SU(3)$ gauge group on such a curve, which has self-intersection $+1$:
for $SU(2)$ there are 22 matter hypermultiplets transforming in the
fundamental representation, and for $SU(3)$ there are 24 fundamental
hypers.  (See, for example, \cite{Johnson-Taylor}, \S2.5;
the fact that only matter hypermultiplets in the fundamental
representations can arise can be proven simply
from the fact that matter representations of $SU(n)$ associated with
Young diagrams with more than one column can only be present when the
$SU(n)$ is realized on a curve of genus $g > 0$ \cite{kpt}.)

We also consider six-dimensional compactifications with base surfaces
that give rise to non-Higgsable $SU(3)$ factors; such factors always
arise on $-3$ curves in the base, as in the Hirzebruch surface $\bF_3$,
and in six dimensions there cannot be any matter charged under the
resulting $SU(3)$.  In six dimensions, there are no non-Higgsable
realizations of $SU(2)$ that are relevant for the gauge groups of
interest here; the only way that a non-Higgsable $SU(2)$ can arise is
in combination with either a $G_2$ or $SO(7)$ gauge group factor
\cite{clusters}.

\subsubsection{$SU(2)$ in 6D models} 

First, we illustrate the difference between tuning an $SU(2)$ on the
divisor $Z = \{z = 0\}$ on $B_2 = \P^2$ with a type $III$ singularity
versus a type $I_2$ singularity.  In local coordinates $z, \yy$, we can
expand
\begin{eqnarray*}
f (z, \yy) & = &  f_0^{(12)} (\yy)+ f_1^{(11)}(\yy)z + \cdots + f_{12}^{(0)} z^{12}\\
g (z, \yy) & = & g_0^{(18)} (\yy) + g_1^{(17)} (\yy) z + \cdots + g_{18}^{(0)}z^{18} \,,
\end{eqnarray*}
where $f, g$ are polynomials in $z, \yy$ of degrees 12, 18 respectively,
since the anti-canonical class of $\P^2$ is $- K = 3H$, with $H$ the
hyperplane class, and $f, g$ are sections of $\cO(-4K)$ and
$\cO(-6K)$, respectively.  To tune an $SU(2)$ on $Z$ with a type $I_2$
singularity, we need to arrange for $\Delta = 4f^{3}+27g^2$ to vanish
to order $z^2$.  We choose $f_0^{(12)}= -3 \phi_{(6)}^2, g_0^{(18)}= 2
\phi_{(6)}^{3}, g_1^{(17)}= - \phi_{(6)}f_1^{(11)}$, and find
\begin{equation}
 \Delta =9\,
z^2 \phi_{(6)}^2 \left( - (f_1^{(11)})^2
 +12\, \phi_{(6)}g_2^{(16)}+ 12 \phi_{(6)}^2 f_2^{(10)}\right) + {\cal O}(z^{3})\,.
\label{eq:i2-locus}
\end{equation}
As discussed, for example, in \cite{mt-singularities}, the 22 points
where the term in parentheses vanishes are locations where the
residual discriminant locus intersects  $Z$, the singularity is
enhanced to $I_3$, and there is a fundamental matter representation of
the $SU(2)$ through the standard Katz-Vafa rank one enhancement
\cite{Katz-Vafa}.  At the points where $\phi_{(6)}$ vanishes, there would
be an antisymmetric representation if the gauge group were $SU(n)$
with $n> 3$ (or an anti-fundamental for $n= 3$), but this
representation is trivial for $n= 2$.

Now, let us consider the tuning of a type $III$ singularity on the
same locus $Z$.  To achieve this we set $f_0^{(12)}= g_0^{(18)}= g_1^{(17)}= 0$.
The discriminant is then
\begin{equation}
 \Delta = 4(f_1^{(11)})^{3} z^2 + {\cal O} (z^{3}) \,.
\end{equation}
We see that at points where $f_1^{(11)}$ vanishes, the type $III$
singularity is enhanced to a type $IV$ (2, 2, 4) singularity.  Since
anomaly cancellation uniquely determines the matter content, which
must consist of 22 matter fields in the fundamental representation of
$SU(2)$, we see that at each such enhancement point there are two
fundamental matter fields.  This multiplicity can also be determined
by an analysis of the singularity structure, as described in
\cite{GrassiMorrisonFirst,GrassiMorrison}.  This shows that when a
codimension one type $III$ singularity is enhanced at a codimension
two locus to a type $IV$ singularity, there is fundamental matter,
with multiplicity two in the case of six-dimensional theories.

\subsubsection{$SU(3)$ in six dimensions} 

A similar argument to the one above shows that while tuning an $SU(3)$ on
an $I_3$ gives 24 fundamental representations at separate points on a
codimension one curve (really 21 fundamentals where there is an
enhancement $I_3 \rightarrow I_4$, and 3 ``antisymmetrics'' where
$\phi_{(6)}=\varphi_{(3)}^2$ vanishes), when the $SU(3)$ is tuned on a type $IV$
singularity, 
we have $f_0^{(12)}=  f_1^{(11)}=g_0^{(18)}= g_1^{(17)}= 0$, and
$g_2^{(16)}= 
(\gamma^{(8)})^2$,
where the condition that $g_2$ is a perfect square is necessary for
the monodromy at the type $IV$ locus to give an $SU(3)$ instead of an
$SU(2)$ gauge group factor \cite{Bershadsky-all, Morrison-sn}.  In this situation,
\begin{equation}
 \Delta = 27(\gamma^{(8)})^4 z^{4} + {\cal  O} (z^{5}) \,,
\end{equation}
and the 24 matter fields arise at eight places where the type $IV$
singularity is enhanced to a $I_0^*$ (2, 3, 6) singularity.  Thus, at
codimension two loci where a type $IV$ singularity giving a $SU(3)$ is
enhanced to a type $I_0^*$ singularity, there is matter in the
fundamental representation of $SU(3)$, with multiplicity 3 in the case
of six-dimensional theories. This result was also derived in
\cite{GrassiMorrison}.

\subsubsection{$SU(3) \times SU(2)$ in six dimensions} 
\label{sec:6-32}

Now, let us consider a situation where  type $III$ and type $IV$
singularities intersect transversely with an $SU(3)$
on the type $IV$.  On $\P^2$ this can be arranged by choosing
\begin{eqnarray*}
f & = & z \yy^2 \tilde{f}\\
g & = &  z^2 \yy^2 (\gamma_{(7)})^2 + \cdots \,
\end{eqnarray*}
which gives
\begin{equation}
 \Delta = z^{3} \yy^{4} \left( 27 z \gamma^{4} +4\yy^2
 \tilde{f}^{3}\cdots \right) \,,
\end{equation}
so the singularity at the intersection point $z = \yy = 0$ is type
$IV^{*}$ (3, 4, 8).  Since $\tilde{f}, \gamma$ have degrees 9, 7
respectively, we see that the type $III$ curve $Z$ is enhanced to type
$IV$ in the fashion described above at 9 points away from the
$IV$-$III$ intersection, giving 18 fundamentals, and the type $IV$
curve $\YY$ is enhanced to type $I_0^*$ at 7 points away from the
intersection, giving 21 fundamentals.  Anomaly cancellation indicates
that there is one bifundamental $({\bf 3}, {\bf 2})$ at the $IV$-$III$
intersection point, and
requires a total of $22$ ${\fund_2}'s$ and $24$ ${\fund_3}'s$. Thus,
anomaly cancellation requires that there is  one $SU(2)$ fundamental
$({\bf 1},
 {\bf 2})$ and one $SU(3)$ fundamental $({\bf 3}, {\bf 1})$ that must also be localized
at the intersection point. 

To summarize, in the 6D model the $SU(3)\times SU(2)$ charged matter
localized at the $IV$-$III$ intersection consists of one
hypermultiplet each in the $({\bf 3}, {\bf 2}), ({\bf 1}, {\bf 2}), ({\bf 3}, {\bf 1})$ representations.
These non-trivial $SU(3)\times SU(2)$ representations match those of
the deformation analysis of the previous section.  Note, as
mentioned earlier, that this anomaly analysis is not sensitive to uncharged
$SU(3)\times SU(2)$ singlets that are localized at the $IV$-$III$(-$I_2$)
collision, which the deformation analysis suggests must exist.

We can use a similar analysis to construct an $SU(3) \times SU(2)$
gauge group through an intersection of type $IV$ and $I_2$
singularities, beginning with the form of the discriminant  (\ref{eq:i2-locus})
given the
$I_2$ locus on $z = 0$, and
where $\yy | \phi, \yy^2 | f_1^{(11)}, \ldots$.  The result is
basically the same as in the preceding example, except that 19
fundamentals arise on the $I_2$ locus away from the $IV$-$I_2$, occurring
at 19 distinct points  each containing a single multiplet, unlike the
type $III$ $SU(2)$
case where the multiplets appear in pairs.  
Localized at the 6D $IV$-$I_2$ intersection
there is one bifundamental $({\bf 3}, {\bf 2})$, and one  fundamental $({\bf 3}, {\bf 1})$,
but no $({\bf 1}, {\bf 2})$.

Finally, we consider the $IV$-$IVm$ intersection.  To understand the
spectrum here, it is simplest to begin with a $IV$-$IV$ intersection
where both divisors carry an $SU(3)$.  In this case, we have $f = z^2
\yy^2 \tilde{f}, g = z^2 \yy^2 (\gamma_{(7)})^2 + \cdots$.  Just as for
the $SU(3)$ of the $IV$-$III$ intersection, each of the $SU(3)$'s has
21 fundamentals localized at 7 points away from the intersection, and
there is a $({\bf 3}, {\bf 3})$ localized at the intersection, which
suffices to saturate the anomalies.  We can now break one of the
$SU(3)$'s by Higgsing two of the fundamentals. 
 Geometrically, this can be done
in a way that corresponds to turning on a nontrivial monodromy for one
of the $SU(3)$ factors; this occurs if we take $g = z^2 \yy^2
(\gamma_{(7)}^2 + \yy\gamma_{13}) + {\cal O}(z^{3})$.  The resulting
$SU(3) \times SU(2)$ has 21 $({\bf 3}, {\bf 1})$'s localized away from the
intersection, 19 $({\bf 1}, {\bf 2})$'s localized away from the intersection, and
one each of $({\bf 3}, {\bf 2}), ({\bf 3}, {\bf 1})$ localized at the intersection.

The detailed counting from the geometry here is rather subtle;
following \cite{Katz-Morrison-Plesser, GrassiMorrison} the $SU(2)$
fundamentals are nonlocal matter on the monodromy cover of the $SU(2)$
curve.  In particular, there are 14 branch points in the monodromy
cover.  The total space is thus a 14 times branched cover of a
rational curve, which has genus 6.  Each branch point contributes a
half hypermultiplet in the fundamental of $SU(2)$, and each
contribution to the genus  gives an additional 2 nonlocal
fundamentals.  This gives a total of $2 \times 6+7 = 19$ fundamentals
of $SU(2)$
that are either nonlocal or localized away from the intersection point, confirming the
preceding Higgsing analysis, and showing that the matter at the
intersection point is simply $({\bf 3}, {\bf 2}) + ({\bf 3}, {\bf 1})$.

\subsubsection{Non-Higgsable $SU(3)$
in six dimensions}

In six dimensions, the complete set of non-Higgsable clusters was
determined in \cite{clusters}.  A non-Higgsable $SU(3)$ arises on a
type $IV$ singularity over any curve in the (two-dimensional) base
that has self-intersection $-3$ and that does not intersect any curve
of self-intersection $-2$ or below.  Anomaly cancellation in the 6D
theory shows that any non-Higgsable $SU(3)$, which must lie on a $-3$
curve, cannot have any associated charged matter fields.  This can
also be seen geometrically, as argued below.  Since a non-Higgsable
$SU(3)$ cannot have any charged matter, it is not possible to have a
non-Higgsable $SU(3)$ intersecting a Higgsable $SU(2)$.  Unlike
for
$SU(3)$, there are no non-Higgsable clusters in six dimensions
that give rise to the gauge groups $SU(2)$ or $SU(3) \times SU(2)$.
The only cases in
which a non-Higgsable $SU(2)$ gauge group factor can arise are in
non-Higgsable clusters that support gauge groups $G_2 \times SU(2)$ or
$SU(2) \times SO(7) \times SU(2)$.  As we show in the next subsections
(\ref{sec:examples.IV}, \ref{sec:examples.IV-III}),
while $SU(3) \times SU(2)$ thus cannot be realized in six dimensions
in such a way that either or both of the factors are non-Higgsable, in
four dimensions a non-Higgsable $SU(3)$ can be combined either with a
Higgsable or non-Higgsable $SU(2)$.

The simplest explicit example of a non-Higgsable $SU(3)$ in a
six-dimensional F-theory model is on the base $\F_3$.  This base has a
$-3$ curve, over which $f, g$ must vanish to degrees $2, 2$
respectively, with only a single monomial in $g_2$, giving a type $IV$
singularity carrying a $SU(3)$ gauge group.  This result is well known
from the early days of F-theory constructions with dual heterotic
models \cite{Morrison-Vafa-II}, and can be obtained in a variety of
ways, each of which generalizes to four dimensional constructions.
The toric approach can be used, as for $\F_{8}$ in
\S\ref{sec:check-clusters}.  

A convenient and general way of determining
non-Higgsable structures in the toric approach uses the language of
toric {\it fans} \cite{Fulton}.  For $\F_m$ the toric diagram consists
of the set of rays $v_1 = (0, 1), v_2 = (1, 0), v_3 = (0, -1), v_4 =
(-1, - m)$; the rays live in an integral lattice, $v_i \in N= \Z^2$,
and each ray corresponds to an effective divisor in the surface, with
$v_3$ corresponding to the curve of self-intersection $-3$.  The set
of monomials in the Weierstrass model is given by the set of points in
the dual lattice $m = (a, b)\in M = N^*$ that satisfy, for $f$ and $g$
respectively, $\langle m, v_i \rangle \geq -4, -6$ for all rays $v_i$.
For a given $m$
in $f, g$, the degree of vanishing of $m$ on the curve $S$ of
self-intersection $-3$ is given by $4+ \langle m, v_3 \rangle,
6+ \langle m, v_3 \rangle$ respectively.   It is
straightforward to check that
there are no monomials in $f$ or $g$ of degrees 0  or 1 on $S$, and only one
monomial in
$g$ (associated with $m = (-4, 4)$) of degree 2, confirming that
there is a type $IV$ singularity giving a gauge group $SU(3)$ over
this curve.  
This approach is explained and applied systematically for toric
F-theory base surfaces in \cite{toric}.

More abstractly, using the language of algebraic geometry as described
in \cite{clusters}, the anti-canonical class of the base $\F_3$ is $-
K = 2 S +5F$, where in toric language $F$ corresponds to the ray
$v_2(\sim v_4)$, and the curves $S, F$ have the intersection properties $S \cdot
S = -3, S \cdot F = 1, F \cdot F = 0$.  Since the irreducible curve
$S$ has negative self-intersection, it is rigid.  Since $-4K \cdot S =
-4$, the divisor $-4K$ must contain 2 copies of $S$, which means that $f$,
as a section of the line bundle ${\cal O}(-4K)$ associated with $-4K$,
must vanish to degree 2 over $S$.  Similarly, $g$ must also vanish on
$S$ to degree 2, giving again the type $IV$ singularity.  We use
variations on each of these approaches in identifying non-Higgsable
clusters in the following 4D examples.  Finally, as mentioned above,
the F-theory model on $\F_3$ has a dual heterotic description
\cite{Morrison-Vafa-I} in terms of a compactification of heterotic
$E_8 \times E_8$ theory on a K3 surface, where the total instanton
number is divided between the two gauge factors as $24 = 15+9$.  On
the side with 9 instantons, the generic bundle has $E_6$ structure, so the
resulting gauge group is a non-Higgsable $SU(3)$ gauge group with no
charged matter. Thus, non-Higgsable clusters in cases with no matter
are familiar structures also from the heterotic point of view.

Note that the absence of matter for any 6D non-Higgsable $SU(3)$ can
be proven generally from the algebraic geometry point of view.  As
shown in \cite{clusters}, a non-Higgsable $SU(3)$ can only arise on a
rational curve $C$ of self-intersection $C\cdot C = -3$.  Since $- K
\cdot C = -1$, $- nK$ must contain $C$ as an irreducible component
$\lceil n/3 \rceil$ times; this is quantified by the \emph{Zariski
  decomposition} over the rationals $- K = C/3+ X$, where the
coefficient of $C$ is determined by the condition that it is the
minimal value possible such that
the residual component $X$ is effective and satisfies $X
\cdot C \geq 0$.  We then have $- 6K = 2C + X_6$, where the residual
component $X_6$ satisfies $X_6 \cdot C =
0,$ and $-12K = 4C + X_{12}, X_{12}\cdot
C = 0$.  The last condition shows that the discriminant locus can
always be written as $\Delta = z^{4}\tilde{\Delta}$, where
$\tilde{\Delta}$ has a vanishing locus that does not intersect $z =
0$; therefore, there is no associated matter.

As we see below, the corresponding constraint is weaker in four
dimensions, where a non-Higgsable $SU(3)$ can have (geometric) charged
matter, and can be realized in combination with a Higgsable or
non-Higgsable intersecting $SU(2)$ factor.

\subsection{4D models with non-Higgsable $SU(3)$ groups} 
\label{sec:examples.IV}

There are a wide range of 4D models with type $IV$ singularities on
some divisor giving a non-Higgsable gauge
group.  The simplest case is an F-theory model on the base $ B_3 = \P^1
\times  \F_3$.   
In this case the toric analysis    is almost precisely identical to
the $\F_3$ case described above.  The 3D toric fan is generated by the
rays
\begin{eqnarray*}
w_i & = & (v_i, 0), \; i \in \{1, \ldots, 4\}\\
w_{5, 6} & = &  (0, 0, \pm 1) \,.
\end{eqnarray*}
To specify a complete 3D toric fan, we must also specify which pairs
of rays span 2D cones in the fan in a way that triangulates $S^2$; we
do that in this case by having a 2D cone  generated
by $\{w_5,w_i\}$ and $\{w_6,w_i\}$ for each $i <5$, in addition to the
2D cones inherited from $\F_3$
associated with the pairs $\{w_i, w_{i +1}\}, \{w_4, w_1\}$.  The conditions on
monomials are as before.  In this case, however, $g_2$ in the
expansion of $g$ around the divisor $D_3$ is a function of the extra
coordinate with 13 independent parameters, so generically is not a
perfect square, so this gives a non-Higgsable $SU(2)$ gauge group.  A
more constrained construction is needed to get the correct monodromy
for an $SU(3)$ on a type $IV$ singularity in 4 dimensions.

A general class of  4D F-theory compactifications with heterotic duals
were considered in \cite{Anderson-WT}, based on manifolds $B_3$ with
the form of a $\P^1$ bundle over a base $B_2$.  When the base $B_2$ is
a generalized del Pezzo surface, there is a smooth heterotic dual on a
Calabi-Yau threefold that is an elliptic fibration over $B_2$.
When the base $B_2$ is toric, then there is a simple toric description
of the threefold,  with the rays
\begin{eqnarray*}
w_i & = &  (v_i, t_i), \; i = 1, \ldots, N \\
w_{\pm} & = &  (0, 0, \pm 1) \,.
\end{eqnarray*}
Here, the  2D rays  $v_i$ give a toric description of $B_2$, there are
2D cones connecting each of $w_{\pm}$ to each of the $w_i$,  in
addition to the full set of 2D cones inherited from $B_2$.  The
integer parameters $t_i$ represent a divisor $T = \sum_{i}t_i D_i$,
which describes the ``twist'' of the $\P^1$ bundle; the divisors
$D_\pm$ (which are the zero and infinity sections of the $\bP^1$ bundle) corresponding to the rays $w_{\pm}$ are sections of the $\P^1$
bundle, with normal bundles within $B_3$ given by $\mp T$.  Note that
two of the parameters $t_i$ are redundant and can be set to 0 by a
linear transformation of the basis of the lattice $N = \Z^3$ that leaves the
third axis unchanged.

The roughly 4000 bases $B_3$ that are $\P^1$ bundles over a toric
generalized del Pezzo $B_2$ were enumerated in \cite{Anderson-WT}, and
the associated non-Higgsable gauge groups were determined.  Of these,
over 100 had a non-Higgsable $SU(3)$ on a type $IV$ singularity, and
no further non-Higgsable gauge group.  Some simple examples are given
by $B_2 =\F_0 = \P^1 \times \P^1$ with a twist $T = 3 S +3F$, and $B_2
= \F_2$ with a twist $T = 3 S +6F$.  (These are the only cases of
non-Higgsable $SU(3)$'s for  the class of
$\P^1$ bundles over Hirzebruch surfaces
$\F_m$,  sometimes denoted by $\bF_{mqr}$).  The
non-Higgsable $SU(3)$ in these cases can be verified by a direct
computation using toric monomials.  More abstractly, as described in
\cite{Anderson-WT}, the coefficients of $f$ in an expansion $f = f_0+
f_1z + \cdots$ around the divisor $D = D_+$ at a local coordinate $z =
0$ are sections of line bundles on $D$ 
\begin{equation}
 f_k \in {\cal  O} (-4K_D -(4 - k) T) \,,
\end{equation}
and similarly for expansions of $g$ and $\Delta$, with 4 replaced by 6
and 12 respectively.  For the Hirzebruch surface $\F_m$, $- K = 2 S +
(2+ m) F$.  Thus, for the 3D base defined as a $\P^1$ bundle over
$\F_0$ with twist $T = 3 (S + F)$, the expansion around $D_+$ has $f_k
\in {\cal O}((3k -4)(S + F))$, and $g_k \in {\cal O}((3k -6)(S + F))$.
There can only be a section when the relevant divisor is effective,
which occurs when the numerical coefficient $3k -4$ or $3k -6$ is
nonnegative.  This shows that $f, g$ vanish to degrees 2, 2 at $D_+$.
Furthermore, since $g_2 \in {\cal O}(0)$ is a constant, it is a
perfect square and the gauge group is $SU(3)$.  A similar argument
gives a type $IV$ $SU(3)$ singularity in the $\F_2$ case described
above.  Note that in both of these cases, $g_2$ and $\Delta_4$ are
sections of the trivial line bundle.  This means that the residual
discriminant locus does not intersect the divisor carrying the $SU(3)$
gauge group, so just as in the 6D cases described in the previous
section, it is not possible to realize an $SU(2)$ on a divisor
intersecting the divisor carrying the type $IV$ singularity, even by
tuning a Higgsable $SU(2)$.  Such a tuning would cause $g_2$ to vanish
identically, pushing the original type $IV$ singularity to at least an
$I_0^*$ singularity that carries a larger gauge group -- at least
$G_2$.

There are, however, many other instances of non-Higgsable $SU(3)$
gauge groups in the list of examples found in \cite{Anderson-WT},
which involve $\P^1$ bundles over bases that are more complicated,
formed by blowing up Hirzebruch surfaces $\F_m$ at one or more points.
In these more complicated constructions there can be curves that carry
matter charged under the non-Higgsable $SU(3)$, and in some cases
additional divisors on which $SU(2)$ factors can be tuned to intersect
with the non-Higgsable $SU(3)$ without increasing the singularity type
of the original type $IV$ singularity locus.
This gives a variety of situations in which we can have a
non-Higgsable $SU(3)$ combined with a Higgsable $SU(2)$ in a 4D
F-theory model.  Two simple examples of this arise by blowing up
points in the bases $\bF_0, \bF_2$ of the two non-Higgsable $SU(3)$
examples described above.

First, consider blowing up $F_0 = \P^1 \times \P^1$ at a point, which
gives the del Pezzo surface $dP_2$.  This surface can be described
torically, and we can construct a 3D base $B_3$ as a $\P^1$ bundle
over this surface, with the toric rays $w_i= (0, 1, 0), (1, 0, 0), (1,
-1, 1) (0, -1, 3), (-1, 0, 3)$.  It is straightforward to check using
explicit monomials that, like the preceding construction using $F_0$,
this 4D F-theory model has a non-Higgsable $SU(3)$ on the divisor
$D_+$, and that furthermore a type $III$ (Higgsable) $SU(2)$ can be
tuned on the divisor $D_3$ (corresponding to the exceptional divisor
of the blowup on $F_0$), without enhancing the type $IV$ singularity
on $D_+$.  This can be seen more abstractly using the divisor
structure of the base.  The anti-canonical class of the new 2D base
$dP_2$ can be written as $- K_2 = 2 S +2F +3E$, where $S, E, F$
correspond to the toric divisors $D_2, D_3, D_4$ in the preceding fan
description.  In this notation, the twist of the $\P^1$ bundle is $T
=3 S
+ 4E +3F$.  
(Note that, as above, there is a two-parameter  family of equivalences
on the twists, corresponding to the two equivalence relations on the
divisors of a 2D toric fan -- often referred to as the Stanley-Reisner
ideal  -- which allows us to eliminate  two of the divisors in the
base $B_2$ in these expressions for $- K_2, T$.)
In the expansion around $D_+$, the Weierstrass coefficient
$g_2$ is now a section of the line bundle ${\cal O}(2E)$.  In terms of
the Zariski decomposition, this means that $- 6K_3= 2 D_+ + X$, where
$X$ can have a nontrivial intersection with $D_+$ along the curve $D_+
\cap E$.  This means that the non-Higgsable $SU(3)$ in this
construction can have geometric charged matter, which is a necessary
condition for the possibility of constructing a Higgsable $SU(2)$ like
the one that can be realized in this geometry.

A second example of a 4D theory with a non-Higgsable $SU(3)$ and a
Higgsable $SU(2)$ can be found in a similar way by modifying the
example with a non-Higgsable $SU(3)$ constructed from a $\P^1$ bundle
over $F_2$.  Blowing up a generic point on the surface $F_2$ gives a
generalized del Pezzo surface with a toric description.   A toric $\P^1$
bundle over this surface can be constructed using the rays
$w_i = (0, 1, 0), (1, 0, 0), (0, -1, 3), (-1, -2, 6), (-1, -1, 4)$.
Again, explicit checking of monomials demonstrates a non-Higgsable
$SU(3)$ on $D_+$, which can be combined with a Higgsable $SU(2)$ on
$D_5$.  In the more abstract language, we have $- K = 2 S +4F +3E$,
where $S, F, E$ correspond to $D_3, D_4, D_5$ ($E$ is the exceptional
divisor in the blown up base surface $B_2$, as before), and $T = 3 S
+6F +4E$.  Just as in the preceding case, $g_2 \in {\cal  O}(2E)$,
giving the possibility of charged matter under the $SU(3)$, and
matching with the existence of the Higgsable $SU(2)$ on the divisor $E
= D_5$.

A variety of further  constructions can be realized in this way on
base surfaces that involve further blowups; we have identified 27
configurations that allow for this possibility  from the toric
constructions enumerated in \cite{Anderson-WT}.

\subsection{4D models with non-Higgsable $SU(3)\times SU(2)$} 
\label{sec:examples.IV-III} 

Models with non-Higgsable product gauge groups such as
$SU(3) \times SU(2)$ arising on intersecting divisors do not appear in
the class of models considered in \cite{Anderson-WT}, since in
F-theory
models
with  smooth heterotic duals gauge groups can only arise on the
sections $D_\pm$ of a $\P^1$ bundle.  By considering a more general
class of bases $B_3$ that are $\P^1$ bundles over the more general
class of toric
surfaces $B_2$ identified in \cite{toric}
that can act as bases for elliptically fibered Calabi-Yau threefolds,
including those bases
that include curves of self-intersection $-3$ or below, we find
models with more general non-Higgsable product group structures.  A
systematic scan over such more general F-theory $\P^1$ bundle
bases (more general results from which will appear elsewhere
\cite{Halverson-WT}) reveals
a number of models with non-Higgsable $SU(3) \times SU(2)$ gauge
groups coming from intersecting $IV$-$III$(-$I_1$) singularities.  An example
of such a model arises for a toric base  constructed by blowing up the
Hirzebruch surface $\F_2$ at two distinct points on the $-2$ curve
$S$.  A $\P^1$ bundle over this base can be constructed using the
toric rays
\begin{eqnarray*}
w_1 & = & (0, 1, 0) \\ 
w_2 & = & (1, 0, 0) \\
w_3 & = & (1, -1, 1)\\
w_4 & = & (0, -1, 2)\\
w_5 & = & (-1, -3, 6) \\ 
w_6 & = & (-1, -2, 5)\\
w_{\pm} & = &  (0, 0, \pm 1) \,,
\end{eqnarray*}
with cone structure as described in the previous subsection.
A direct numerical computation using toric monomials confirms that
this elliptically fibered Calabi-Yau fourfold has type $IV$ and type
$III$ singularities on the divisors $D_4, D_+$, giving rise to a
non-Higgsable gauge group $SU(3) \times SU(2)$.  This can also be
confirmed using a more abstract analysis on the surfaces and curves
analogous to the method used above for analyzing the components of $f,
g$ around a fixed divisor.  A more general description of this
approach will appear in
\cite{mt-4D-clusters}; in this case the analysis is a simple
generalization of what is used above.  For a local analysis around
$D_+$, we have $ -K_+ = D_3+2 D_4+5 D_5+4 D_6$, with a twist
(corresponding to the normal bundle of $D_+$) of $T_+ = D_3+2 D_4+6
D_5+5 D_6$.   So, for example, $g_1$ in an expansion around  $D_+$ is
a section of the line bundle associated with the divisor $-6K_+ -5T_+
= D_3+2 D_4 - D_6$, which is not effective, so $g$  vanishes to degree
at least 2 on $D_+$. Checking the other cases shows that we have a
type $III$ singularity on $D_+$.  The divisor $D_4$ can be seen from
the  local toric structure to have the geometry of a Hirzebruch
surface $\F_1$, with anti-canonical class  $- K_4 =2 D_- + 3 D_5$, and
normal bundle corresponding to a ``twist'' $T = 4 D_5$.  Analyzing the
local structure of $f, g$ with this data gives a type $IV$ (2, 2, 4)
singularity without monodromy, confirming the $SU(3) \times SU(2)$ non-Higgsable cluster
in this geometry.

\sskip

We conclude the set of examples with  an example exhibiting a
non-Higgsable $SU(3) \times SU(2)$ of type $IV$-$IVm$, that is where
the two gauge factors are realized on intersecting type $IV$ divisors,
one without monodromy and one with.  The base is found by blowing up
$\F_0$
(or $\F_1 =dP_1$) at a point giving a $dP_2$, and then blowing up at
the two
intersection points between pairs of the three -1 curves in $dP_2$.
The resulting base is toric, with a  cyclic
sequence of pairwise intersecting effective toric
divisors of self-intersections $0, -2, -1, -3, -1, -2, 0,$ corresponding
to the rays in the base geometry.  With an appropriate twist we form a
$\P^1$ bundle over this base with rays
\begin{eqnarray*}
w_1 & = & (0, 1, 0) \\ 
w_2 & = & (1, 0, 0) \\
w_3 & = & (2, -1, 0)\\
w_4 & = & (1, -1, 1)\\
w_5 & = & (1, -2, 3) \\ 
w_6 & = & (0, -1, 3)\\
w_7 & = & (-1, 0, 3)\\
w_{\pm} & = &  (0, 0, \pm 1) \,.
\end{eqnarray*}
Direct computation using toric monomials verifies that there are type
$IV$ singularities on the divisors $D_+, D_4$,
with monodromy in the latter case so the gauge group is $SU(3) \times SU(2)$.

\section{Directions for further
  development}
\label{sec:issues}

In this paper we have primarily focused on the geometric structure of
non-Higgsable nonabelian gauge groups.  We have identified the various
ways in which the nonabelian part of the standard model gauge group
$SU(3) \times SU(2)$ can arise in F-theory compactifications to four-
and six-dimensional supergravity theories.  To find a realistic way of
realizing the full standard model in F-theory using a non-Higgsable
$SU(3)$, a number of further aspects of these scenarios must be
analyzed in more detail.  In particular, at least one $U(1)$ factor
must be included in the gauge group, G-flux must be chosen so as not
to alter the gauge group realized at the geometric level, and
the details of the matter content (including multiplicities and
chirality) must be worked out. In this section we give a brief
discussion of the issues involved in these more detailed aspects of a
complete F-theory description of the standard model of particle physics.

We would like to comment that although a careful analysis of $U(1)$
factors and chiral matter will be necessary to obtain a realistic
model in the context of the non-Higgsable $SU(3)$ gauge groups that we
are considering here, recent works \cite{LinWeigand, AllToricHypFibs}
 have given constructions of a chiral spectrum and weak
hypercharge in non-GUT realizations of the standard model in F-theory.
Though these models are Higgsable and the geometries are
different, the results of those works suggest that similar chirality
and hypercharge  structures may also arise in our scenario.

\subsection{Abelian factors}
\label{sec:abelian}

To obtain a realistic model containing (or exactly realizing) the
standard model of particle physics, at least one $U(1)$ factor must
appear in the gauge group.  Abelian gauge
group factors are difficult to study in F-theory as they depend upon
global features of the compactification; nonetheless, there has been
significant recent progress in understanding these factors in the
F-theory context \cite{Grimm-Weigand,Park-WT,Park,Morrison-Park,GfluxAnomalies,Mayrhofer:2012zy,Braun:2013yti,Borchmann:2013jwa,Cvetic-Klevers-1,Borchmann:2013hta,Cvetic-Klevers-2,Cvetic-gkp,Braun-fate,DPS,mt-sections}.  While it is possible to have
non-Higgsable $U(1)$ factors arising on manifolds that have a nonzero
Mordell-Weil rank at all points in their moduli space, such as over
certain bases with a $\C^*$-structure (``semi-toric'' bases) that give
a threefold related to the Schoen construction \cite{Schoen, Martini-WT, Morrison-Park-WT}, these occur only on very special
bases and have not yet been studied in 4 dimensional theories; we
focus attention here on more generic Higgsable $U(1)$ factors that can
be tuned on most F-theory bases.

The main conclusion that we describe here is that while
when $SU(3) \times SU(2)$ is realized in a Higgsable fashion the
particles charged under an additional $U(1)$ will not in general be
localized at the intersection between the divisors carrying the
nonabelian factors without special tuning (even for a $IV$-$III$
intersection), when one of the nonabelian factors is non-Higgsable, in
general there will always be $U(1)$ charged matter localized at the
intersection point.

It was shown in
\cite{Morrison-Park} that any Weierstrass model that admits a $U(1)$
factor can be written in the form
\begin{eqnarray}
\lefteqn{ y^2 = x^3+ (e_1e_3-b^2e_0- \frac{1}{3}e_2^2 ) x
}\label{eq:two} \\ & 
 &
\hspace*{0.4in} +(-e_0e_3^2 +\frac{1}{3}e_1e_2e_3 - \frac{2}{27}e_2^3 + \frac{2}{3}b^2e_0e_2
 -\frac{1}{4}b^2e_1^2) \,,
\nonumber
\end{eqnarray}
where, as described further in \cite{mt-sections}, $b$ is a section of
a line bundle associated with an effective divisor $X$ ($[b]= X$), and
$[e_i] = (i - 4) K + (i - 2) X$.  By making a choice of $X$, we can
tune a model with a $U(1)$ factor and various other desired
singularities by appropriately choosing the sections $b, e_i$.  By
tuning $b \rightarrow 0$ in any such model with a $U(1)$ factor, the
$U(1)$ is enhanced to a nonabelian gauge factor associated with the
divisor $e_3$.  While in some cases the resulting model reaches a
transition point associated with a $(4, 6)$ singularity on a
codimension one locus and is best treated as a superconformal field
theory, the associated nonabelian model can often give a simple
picture of the matter spectrum of the abelian theory.  In particular,
the matter charged under the resulting $U(1)$ is localized on the
vanishing locus of $e_3$, and the charge can often be understood via adjoint
Higgsing of an associated non-abelian model.

As a simple example, consider again a 6D theory on the base $\P^2$,
where the gauge group $SU(3) \times SU(2)$ is realized through a
(tuned, Higgsable) $IV$-$III$ intersection, with the gauge group
factors on the loci $\YY = \{\yy = 0\}, Z = \{z = 0\}$, as described
in \S\ref{sec:6-32}.  The simplest choice of effective divisor for $X$
is the trivial class, so that $b$ is simply a constant.  In this case,
$[e_3]= - K$, so $e_3$ is a cubic.  Tuning an $SU(2)$ on a cubic, from
anomaly cancellation there are 54 fundamentals under the $SU(2)$ and
one adjoint.  Since generically a cubic crosses a line at three
points, there are generically three matter multiplets in each of the
$(3, 1, 2)$ and $(1, 2, 2)$ representations arising from the three
intersection points of each type.  Higgsing the adjoint breaks the
$SU(2)$, giving a $U(1)$ theory with 54 pairs of $+1, -1$ charged
matter fields, 15 of which carry charges under the remaining
nonabelian factors, but none of which need to reside at the $IV$-$III$
intersection point, unless a special cubic is chosen that passes
through that point.  Indeed, 108 is the minimal number of charged
matter fields that are compatible with anomaly cancellation for a
theory with a single $U(1)$ on $\P^2$ \cite{Park-WT}.  This
illustrates the fact that in general when the $SU(3)$ and $SU(2)$ are
Higgsable, matter charged under an additional $U(1)$ need not reside
at the intersection of the divisors carrying the nonabelian factors,
so in particular matter charged under both the $SU(3)$ and $SU(2)$
factors will in general not carry a $U(1)$ charge.  A similar story
holds for 4D models with a Higgsable $SU(3) \times SU(2)$.

When the $SU(3)$ is non-Higgsable and there is a $U(1)$, however,
there will generally be $U(1)$ charged matter at the intersection of
the $SU(3)$ and any nonabelian $SU(2)$ factor.  This follows from an
analysis similar to that carried out in \cite{mt-sections}, which
holds for F-theory models in four dimensions as well as in six.
Basically, the idea is that if there is a non-Higgsable $SU(3)$
supported on a divisor $D$ then $D$ appears as an irreducible
component of $- K$ and can therefore appear as an irreducible
component of $e_3 = - K + X$.  Since matter charged under the $U(1)$
lies at the intersections of $e_3$ with other divisors carrying
nonabelian gauge groups, there will then necessarily be matter charged
under the $U(1)$ at the $SU(3) \times SU(2)$ intersection.  As an
example, analogous to the 6D example described above, a simple way to
arrange for a $U(1)$ factor in the presence of a non-Higgsable type
$IV$ singularity carrying an $SU(3)$ along $\{\yy = 0\}$ that
intersects with an $SU(2)$ on $\{z = 0\}$ is to set $X = [b] = 0 $, so
$b$ is simply a constant.  Since $e_3, e_2, e_1$ are then in the
divisor classes of $-K, -2K, -3K$, each must vanish to order at least
one in $\yy$.  We see then that in the $b \rightarrow 0$ limit, $g$
vanishes to order 3, so we have an $I_0^*$ Kodaira type on $\yy = 0$,
generally associated with a rank enhancement.  In principle, the
charged matter content under the $U(1)$ can be computed in any such
model.  The details will depend on the choice of $X$, but anytime
there is a rank enhancement on the locus $\yy = 0$ there will be
additional charged matter at the intersection point that will acquire
$U(1)$ charge under the Higgsing corresponding to $b \neq 0$.  We
leave the details of such calculations for future work. Note, however,
that if the $SU(2)$ is also non-Higgsable, then in the case $X = 0$,
$e_3, e_2, e_1$ must also vanish on $\{z = 0\}$, which would give a
$(4, 6)$ singularity at the intersection point $\yy = z = 0$ in the
limit $b \rightarrow 0$, corresponding to similar situations in
\cite{mt-sections} where the $U(1)$ cannot be unhiggsed to a
nonabelian factor without going to a superconformal point.  Such a
situation may be avoided by choosing $X$ to be effective and
sufficiently large to reduce the vanishing of $e_3$ on $\yy, z$.

A particularly simple example of the preceding construction for a
model with a $U(1)$ factor can be realized by taking $X$ to be the
trivial class, so $b$ is a constant, and furthermore setting $e_2 =
e_3= 0$.  This can be done on any base, since $X = 0$ is always an
effective divisor.  In this case, $e_0, e_1$ must contain $Z, \YY$ as
components with minimal multiplicities giving the Weierstrass model
\begin{equation}
 y^2 = x^{3} - b^2 z \yy^2 \tilde{e_0} x
- \frac{1}{4} b^2 z^2 \yy^2 \tilde{e_1}^2 \,.
\end{equation}
This model has the type $IV$-$III$ intersection and a U(1) factor.

Finally, we comment briefly on charges.  To get the precise $U(1)$
hypercharges of the standard model, it is necessary to have different
$U(1)$ charges associated with matter in different representations of
the nonabelian part of the gauge group.  Such different charges can
arise when $e_3$ contains one of the loci supporting the nonabelian
gauge factors as a component.  Higher charges can also arise when
$e_3$ is a singular divisor.  As discussed in \cite{mt-sections, kpt},
higher symmetric representations of $SU(2)$ can likely be tuned on
sufficiently singular divisors in the base.  When $e_3$ is such a
divisor, we can locally view the $U(1)$ as a broken nonabelian group,
and higher symmetric representations of $SU(2)$ will give rise to
fields with larger charges under $U(1)$.  We leave a full analysis of
different realizations of $U(1)$ and matter charges in models with
non-Higgsable QCD to further work, though we present a four-dimensional
anomaly analysis in section \ref{sec:discussion},  which shows that  the
minimal
(in
a certain sense) chiral $SU(3)\times SU(2)\times U(1)$ model
contains families of standard model fermions.

\subsection{The the low-energy physical theory and G-flux}
\label{sec:issues-G-flux}

As mentioned earlier, unlike in six-dimensional theories, for
four-dimensional F-theory models the gauge group and matter content do
not necessarily match precisely with those determined
from the geometry.  While in some cases the low-energy physics is not
significantly modified from that described by the geometry, in other
cases G-flux (or possible other mechanisms such as D3-branes) can
modify the spectrum of the theory.  We do not attempt a complete
analysis of these issues here but make a few comments that may help in
framing future work in this direction.  A good review of some of the
general issues associated with G-flux is given in
\cite{Denef-F-theory}.

From the point of view of F-theory as a limit of M-theory, G-flux
corresponds to the four-form flux of the M-theory 3-form potential
wrapped on nontrivial cycles in the (resolved) elliptically fibered Calabi-Yau
fourfold $X$ on which M-theory is compactified.  From the type IIB
point of view, such fluxes can correspond either to three-form flux on
the compactification space or fluxes in the world-volume of
seven-branes. G-flux must satisfy several conditions.  First, tadpole
cancellation dictates that
\begin{equation}
\chi (X)/24 = \int G \wedge G + N_{{\rm D3}} \,,
\label{eq:Euler-flux} 
\end{equation}
where $\chi (X)$ is the Euler character of the fourfold $X$, and
$N_{{\rm D3}}$ is the number of D3-branes in the system.
G-flux must also satisfy the parity condition \cite{Witten-flux} 
\begin{equation}
G + \frac{c_2}{2} \in H^4(X,\bZ)
\label{eq:G-parity-condition}
\end{equation}
These conditions generally require nonzero G-flux, though
if $c_2/2\in H^4(X,\bZ)$ it is possible to take $G=0$, satisfying
(\ref{eq:Euler-flux}) entirely by the inclusion of
D3-branes. In general, for a given geometry the number of possible
G-flux configurations that satisfy the conditions
(\ref{eq:Euler-flux}) and (\ref{eq:G-parity-condition}) is extremely
large, which is famously helpful for the cosmological constant problem
but problematic for any attempts to explicitly compute solutions on a
case-by-case basis.

When G-flux is present, several effects can in principle change the
spectrum of the theory, modifying the gauge group and matter content
from that of a particular member in the family of Weierstrass models.  The
first issue is the superpotential $W = \int G \wedge \Omega$
\cite{gvw}.  In general, this superpotential stabilizes many of the
moduli.  If the moduli are stabilized at a generic point in the
complex structure moduli space, then the gauge group is that described
by the non-Higgsable clusters in the base.  It is possible, however,
that in some situations the superpotential may stabilize the moduli on
a locus in moduli space with an enhanced gauge symmetry, so that the
resulting 4D theory has a larger gauge symmetry than is indicated by
the structure of non-Higgsable clusters.  It seems plausible that for
many choices of G-flux (and perhaps most or all) no such symmetry
enhancement would occur, but it is a logical possibility that such a
symmetry enhancement can be forced in certain situations.

A second way in which the gauge group can be modified is by turning on
fluxes on the seven-brane world volume in the F-theory picture.  Such
fluxes can break the symmetry group supported on the seven-branes in
question.  While some specific examples have been identified where
this occurs (see {\it e.g.} \cite{Anderson-WT} and references
therein), it is not clear how generally (given all possible flux
choices) this can or must occur in models.  From a practical point of
view, however, constraints on the flux quanta that are required by the
absence of flux breaking are to some extent understood; see {\it e.g.}
\cite{GfluxResolved}.  In many known cases, particularly those with
heterotic duals, solutions exist, and therefore there exist consistent
choices of G-flux which do not break the geometrically-determined
gauge group.

Finally, G-flux can affect the matter spectrum of the theory.  In the
absence of G-flux, the matter spectrum is purely non-chiral, $``{\cal
  N}= 2"$ like matter.  When G-flux is included, the matter spectrum
generally develops a chiral component.  In particular, the chiral index
of states in a four-dimensional $\cN=1$ F-theory compactification can
be determined by the choice of $G$-flux in the related
three-dimensional $\cN=2$ M-theory compactification. Over the last few
years this relationship has been explored in great detail; see for
example \cite{Marsano-flux, Blumenhagen-F-theory} for analyses of
G-flux in global models in the context of spectral covers,
\cite{GfluxResolved,Grimm-Hayashi} for analyses based on the resolved geometry and
M-theory, and
\cite{GfluxAnomalies} for an exploration of the relationship between
this flux and anomaly cancellation in the four-dimensional F-theory
compactification.

Understanding how these various aspects of G-flux interact with
non-Higgsable structures in an F-theory compactification presents a
variety of
questions for further work.  In the most straightforward scenario, to
realize a realistic standard model using a non-Higgsable $SU(3)$, we
would want to identify an elliptically fibered CY fourfold with a
non-Higgsable $SU(3)$, and either a non-Higgsable or Higgsable
$SU(2)$, in a setup where G-flux neither enhanced or diminished the
geometric F-theory gauge group (as often assumed in the F-theory
model building literature), but where the G-flux would give rise
to the proper chiral spectrum for the standard model.  The explicit
examples given in section \ref{sec:examples} could serve as a starting
point for such an analysis.  

As described in the next section, four-dimensional anomaly
cancellation places fairly strong constraints on the resulting
spectrum, so that multiple generations of standard model matter is one
of the few simple solutions that is consistent with all known
conditions.  There are other possible scenarios, however, as well, in
which non-Higgsable structures could play a role in a realistic
F-theory construction of the standard model.  For example, there could
be a non-Higgsable $E_6$ GUT group that is broken by flux on the
seven-branes that carry the group.  Or, moduli stabilization from the
F-theory superpotential could push the theory to a locus with an
unbroken gauge group larger than the geometric non-Higgsable gauge
group.  We leave further investigation of these interesting questions
to future work.

\sskip

It is clear any case, however, that an
important issue that must be understood more clearly to progress
with the ideas that we have developed in this paper is the
relationship between configurations in 4D F-theory geometry that are
non-Higgsable from the point of view of complex structure moduli space
and the physics of the corresponding low-energy 4D supergravity theories.
In six-dimensional theories, gauge groups and matter that are
geometrically non-Higgsable correspond to gauge groups and matter in
the low-energy theory that are physically non-Higgsable.  In these
cases, the non-Higgsability of the theory can be understood from the
point of view of low-energy field theory.  For example,
the matter in a 6D theory with a non-Higgsable $G_2 \times SU(2)$
cluster lives in a single half-hypermultiplet of the $SU(2)$ (times a
${\bf 7}+ {\bf 1}$ of the $G_2$), and cannot be Higgsed because, from
the ${\cal N}= 1$ point of view, the D-term constraints cannot cancel
for a single fundamental of $SU(2)$.  

On the other hand, in many 4D models with geometrically non-Higgsable
structure such as we have considered here, the associated matter
fields appear to allow for D-flat symmetry breaking directions that
may or may not be F-flat.  In the framework of $d=4$ $\cN=1$
Lagrangian supergravity, the absence of a superpotential which
obstructs such a flat direction would correspond to a Higgsable
theory; geometric non-Higgsability may provide evidence that such a
superpotential must exist.  As a concrete example, in a theory with
fields $Q, D, L$ transforming in the representations $({\bf 3}, {\bf
  2}), ({\bf \bar{3}}, {\bf 1}), ({\bf 1}, {\bf 2})$ of a group $SU(3)
\times SU(2)$, the scalar combination $Q D L$ is gauge invariant and
can in principle be turned on to Higgs the gauge group.  In a theory
where the group is geometrically non-Higgsable, these fields do not
correspond to complex structure moduli of the corresponding elliptic
fibration.  The simplest interpretation of the geometric
non-Higgsability would be that while the fields $Q, D, L$ may be
massless at quadratic order, a cubic term $QDL$ in the superpotential
may stabilize these fields at quartic order.  Indeed, in general
without some explicit symmetry (such as R-parity) to forbid such
terms, they are expected to arise from perturbative and/or
nonperturbative effects.  Understanding how this story works out in 4D
F-theory constructions is an important question that we hope will be
addressed by further work on these models, or more broadly on
non-Higgsable clusters.

While the simplest and most satisfactory scenario might be that in
those cases where the geometrically non-Higgsable gauge group and
matter content persist in the low-energy 4D theory the
non-Higgsability of the matter fields automatically arises from D-term
and/or F-term constraints in the supergravity theory, we cannot rule
out completely other possibilities.  For example, there are additional
degrees of freedom other than the complex structure moduli that are
relevant in the low-energy 4D theory, but not yet fully understood in the
context of F-theory. It may be possible that these degrees of freedom
can give expectation values to matter fields which Higgs the gauge
theory, even if complex structure deformations may not.  Such degrees
of freedom include continuous moduli associated with world volume
gauge fields on the seven-branes in the IIB picture, or associated
with $h^{2, 1}$ of the Calabi-Yau fourfold in the M-theory picture.

At this point in time, F-theory is not a completely defined framework,
and there is not yet a systematic way of constructing the complete
low-energy supergravity theory associated with a given F-theory
compactification.  The configurations that contain non-Higgsable
clusters necessarily involve nonperturbative physics from the string
theory point of view, so new insights may be needed to clarify the
physical mechanisms involved in vacuum solutions constructed using
these geometries. Nevertheless, they are an interesting
  direction for current and future research.

\section{Summary and discussion}
\label{sec:discussion}

In this final section, we summarize the main results of the paper,
make a brief but potentially significant argument from four-dimensional anomaly 
cancellation, and
discuss some possible physical lessons. 

\subsection{A Summary of Non-Higgsable QCD}
\label{sec:threerealizations}
\label{sec:summary.spectra}

In this section we review some of the main results from sections
\ref{sec:nonhiggsableQCD} and \ref{sec:examples} on the three
qualitatively different realizations of non-Higgsable QCD and their
associated non-abelian matter spectra; see Table
\ref{tab:SU3SU2spectrum}.  In section \ref{sec:SM} we gave an overview
of the possible ways in which the standard model could be realized in
an F-theory compactification, emphasizing the possible role of
non-Higgsable clusters.  For much of the rest of the paper we have
focused on ``non-Higgsable QCD'' in which the $SU(3)$ factor in the
standard model gauge group is a geometrically non-Higgsable gauge
factor realized on a Kodaira type $IV$ codimension one singularity in
the F-theory base.

We have shown that there are three qualitatively different possible
realizations of non-Higgsable QCD, depending upon how the $SU(2)$
factor of the standard model gauge group is realized.  The three
possibilities for the $SU(2)$ are from seven-branes associated to a
type $IV$ fiber with outer monodromy, a type $III$ fiber, or an $I_2$
fiber.  We refer to these scenarios as $IV$-$IVm$, $IV$-$III$, and
$IV$-$I_2$, respectively.  In each case we have analyzed the geometric
matter content along the curve in $B_3$ lying at the intersection of
the divisors supporting the $SU(3)$ and $SU(2)$ gauge group factors.
This determines the Lie algebra representations, but gives matter that
is
non-chiral in the absence of G-flux; as such, G-flux must be
incorporated to produce chiral matter as described in section
\ref{sec:issues-G-flux}.  The multiplicity (number of generations) of
matter will also be a model-dependent quantity that must be analyzed
separately in specific examples.  Here we briefly summarize the Lie
algebraic (geometric) structure of matter in each of the three
scenarios.  In each case we have identified in section
\ref{sec:examples} some specific examples of (toric) threefold bases
$B_3$ in which the different types of non-Higgsable and Higgsable
$SU(3)$ and $SU(2)$ group factors are realized.  These examples serve
as existence proofs for the general ideas described in the paper, and
as useful starting points for further analysis.

\sskip
\noindent\emph{The $IV$-$III$ Case}

The possibility of $IV$-$III$ non-Higgsable QCD was introduced in
section \ref{sec:nonhiggsableQCD}, including a string junction
analysis of two-cycles in an elliptic surface near the $IV$-$III$
intersection. One interesting feature is that an extra brane which
carries an $I_1$ singularity is always present at this intersection, and
string junctions may end on this brane. Together with the anomaly analysis of section
\ref{sec:examples}, we found that
the spectrum of matter  representations
on the $IV$-$III$
intersection curve that transform in  nontrivial $SU(3)\times SU(2)$ representations 
is (denoting the fundamental of $SU(N)$ as $\fund_N$),
\begin{equation}
(\fund_3,\fund_2) + \fund_3 + \fund_2
\end{equation}
which is precisely the set of non-trivial $SU(3)\times SU(2)$
representations that are realized by matter fields in the standard
model (up to conjugates, which are automatically included at the
geometric level).  The string junction analysis also showed
the existence of
$SU(3)\times SU(2)$ singlet states on the same matter curve; such
singlets embed into the intermediate $16$ of $SO(10)$ (realized on the
extra brane) as expected of right-handed electrons or neutrinos.  As
discussed in section \ref{sec:issues}, when an extra $U(1)$ factor is
incorporated in a model of this type,  matter at the intersection
point acquires charges under the $U(1)$, and the presence of G-flux
produces a chiral matter spectrum.  As we will show shortly, anomaly
cancellation then strongly restricts the set of possible
multiplicities and $U(1)$ charges, so that in a simple picture where
there are multiple generations of a common matter structure, the set
of fields in the standard model is one of the only possible consistent
solutions.

\begin{table}[t]
  \centering
  \begin{tabular}{cc}
    \vspace{.1cm}Realization & \shs \shs \shs \shs Non-trivial $SU(3)\times SU(2)$ Spectrum \shs \shs \shs \shs \\ \hline\hline
    $IV$-$III$ & $(\fund_3,\fund_2) + \fund_3 + \fund_2$ \\
    $IV$-$IVm$ & $(\fund_3,\fund_2) + \fund_3$  \\
    $IV$-$I_2$ & $(\fund_3,\fund_2) + \fund_3$ \\ \hline\hline
  \end{tabular}
  \caption{The non-trivial $SU(3)\times SU(2)$ representations at the collision of $SU(3)$ and $SU(2)$ seven-branes for each of the three realizations of non-Higgsable QCD. }
  \label{tab:SU3SU2spectrum}
\end{table}

\sskip
\noindent\emph{The $IV$-$IVm$ and $IV$-$I_2$ Cases}

In the remaining cases, the $SU(2)$ can be Higgsable or non-Higgsable
when realized on a type $IV$ singularity with monodromy, and must be
Higgsable when realized on a type $I_2$ singularity.
In section
\ref{sec:examples} we performed an anomaly analysis  similar to the
one for the $IV$-$III$
case in order to determine the $SU(3)\times SU(2)$ matter
representations at the associated seven-brane collisions. In both
cases we have found that the non-trivial $SU(3)\times SU(2)$ spectrum at the
collision of the $SU(3)$ and $SU(2)$ seven-branes is
\begin{equation}
(\fund_3,\fund_2) +\fund_3.
\end{equation}
Notably, the $\fund_2$ that is present at the $IV$-$III$ collision is
not present at the $IV$-$IVm$ or $IV$-$I_2$ collision.  This makes
these scenarios potentially less attractive for realizing the full
standard model spectrum without incorporating some additional
structure.

\subsection{The Minimal Chiral $SU(3)\times SU(2)\times U(1)$ Model Gives
Standard Model Generations}
\label{sec:minimal chiral}

Anomaly cancellation in four dimensions constrains the possible chiral
models that one might obtain from non-Higgsable QCD. We briefly
discuss this here, since it demonstrates the plausibility of obtaining
chiral standard models in concrete four-dimensional non-Higgsable QCD
models with G-flux.  This analysis is similar to well-known results in
the field theory literature, though presented here from the point of
view of non-Higgsable QCD.

From the three geometric realizations of non-Higgsable QCD, we know
that the allowed matter representations are $({\bf 3}, {\bf 2}), ({\bf
  3}, {\bf 1}), ({\bf 1}, {\bf 2})$, the conjugates of these
representations, and the singlet representation.  In a chiral
four-dimensional model, $SU(3)$ anomaly cancellation implies that the
number of ${\bf 3}$ and ${\bf \bar{3}}$ fields must match, so that if
we have $N$ chiral $({\bf 3}, {\bf 2})$ matter fields, we must
have\footnote{In many cases it is known \cite{GfluxAnomalies} that a
  consistent choice of G-flux automatically ensures anomaly
  cancellation.} a corresponding $2N$ $({\bf \bar{3}}, {\bf 1})$
fields, suggesting $N$ generations each with one $({\bf 3}, {\bf 2})$
and two copies of $({\bf 3}, {\bf 2})$. The number $K$ of $({\bf
  1,2})$ chiral fermions (multiplets) is not as precisely fixed since
cubic $SU(2)$ non-abelian anomaly cancellation is trivially
satisfied\footnote{Though in some contexts in the landscape, there are
  additional constraints on $SU(2)$ theories related to brane
  nucleation \cite{AnomalyNucleation}.}.  In the absence of additional
chiral $SU(2)$ fermions, however, the global $SU(2)$ anomaly
\cite{WittenAnomaly} requires $(3N+K)\equiv 0\, (\text{mod 2})$.  We
take as an assumption, though it is perhaps also natural to expect,
that these $({\bf 1}, {\bf 2})$ multiplets (if they exist) might arise
in a family with the other fields. That is, we take $K=P\, N$ where
$P$ is constrained by the global anomaly depending on $N$. This
yields\footnote{We emphasize that this is a statement about chiral
  matter; in some regions of moduli space there may be additional
  vector pairs, which could be interpreted as Higgs doublets.} $N$
copies of $({\bf {3}}, {\bf 2})+2\times ({\bf \bar{3}}, {\bf 1})+P
\times \, ({\bf {1}}, {\bf 2})+N_S \times ({\bf 1,1})$, where $N_S$ is
the number of $SU(3)\times SU(2)$ singlets per family.

Given this structure, we would like to determine the minimal
consistent chiral $SU(3)\times SU(2)\times U(1)$ model, as determined
by the minimal $P$ and $N_S$. For $P=0$, the only anomaly free
$U(1)$ is the non-chiral $U(1)$ that assigns opposite charges to the
two copies of right-handed quarks; so to have a chiral model we must
proceed to $P=1$. For $P=1$ and $N_S=0$, the only anomaly free $U(1)$
is the same non-chiral $U(1)$ just mentioned; this exists for any
value of $N_S$. At $P=N_S=1$, however, there is another anomaly free
$U(1)$. It is chiral and has charges (denoted as subscripts)
\begin{equation}
({\bf {3}}, {\bf 2})_1+({\bf
  \bar{3}}, {\bf 1})_{-4}+ ({\bf
  \bar{3}}, {\bf 1})_{2}+({\bf {1}}, {\bf 2})_{-3}+({\bf 1,1})_6.
\end{equation}
This is precisely a generation of standard model fermions; the minimal
chiral $U(1)$ is the weak hypercharge. In summary, we see that the
minimal chiral $SU(3)\times SU(2) \times U(1)$ model that may arise
from non-Higgsable QCD gives rise to families of standard model
fermions.

\subsection{Discussion}

We have found that incorporating a non-Higgsable $SU(3)$ or $SU(3)
\times SU(2)$ into an F-theory compactification leads to a scenario in
which certain features of the standard model, such as the unbroken QCD
sector and the  standard model matter spectrum, seem very
natural.  It is interesting and perhaps suggestive that $SU(3)$,
$SU(2)$, and $SU(3) \times SU(2)$ can be realized in a non-Higgsable
fashion in F-theory, while $SU(n)$ for $n> 3$ cannot be realized in
this way.  We conclude this paper with some speculations regarding the
physical implications of such a scenario, and also some comparison to
related ideas in the literature.

\sskip 
\noindent \emph{Enhanced Symmetry Points in the Landscape}
\vspace{.1cm}

Scenarios \cite{KKLT,LVS} for moduli stabilization in type IIB string
theory have led to a general consensus that there should exist a large landscape
of metastable string vacua \cite{landscape-Susskind,
Grana, Douglas-Kachru}.  The existence of this
large set of possible vacuum solutions prompts a number of interesting
physical questions -- for example, whether there exist identifiable vacua
with the same properties as those we observe in nature, or whether
there exists a dynamical mechanism for vacuum selection.

A related question is why, in a theory 
that can be characterized in terms of a large space of possible
scalar field values with many metastable vacua distributed throughout
it, one might expect stabilization to occur at a point with enhanced
symmetry. More specifically, why would the scalar potential conspire
to give rise to metastable vacua at enhanced symmetry points, and what
dynamical mechanism might drive the moduli toward the enhanced
symmetry point?  This question is interesting because of the
symmetries or near-symmetries observed in our vacuum, and also the
common expectation that a generic metastable vacuum in the landscape
is \emph{not} at an enhanced symmetry point. A number of works have studied
this issue and demonstrated the existence of mechanisms that can
trap
moduli at enhanced symmetry points; see for example \cite{Beauty} and
references therein, and also \cite{ESP-I,ESP-II}.

One of the messages of this paper is that non-Higgsable clusters could
provide an alternative answer to the same question. As opposed to
explaining why moduli stabilization selects a vacuum at a
\emph{special} point in moduli space with enhanced symmetry, as in
other approaches, non-Higgsable clusters circumvent the problem
completely, since for geometries with one or more non-Higgsable
clusters, a \emph{generic} point in the moduli space has enhanced
symmetry.  Moduli stabilization must still occur, of course, but
stabilization at a generic point is sufficient to obtain enhanced
symmetry.

  Based
on what is currently known about non-Higgsable clusters and the
structure of elliptically fibered Calabi-Yau manifolds, we believe
that a common assumption about the landscape must be revisited. The
assumption, used for example in the first approach described above, is
that vacua with enhanced symmetry or light particles comprise a very
small subset of the space of all metastable vacua.  This assumption
seems to be valid only to the extent that non-Higgsable clusters are
non-generic in the landscape.  F-theory, however, seems to provide at
this point the largest range of possible string theory vacuum
solutions,  and seems to generically produce
vacua containing non-Higgsable clusters.

From what is currently understood, in fact, it appears that the
overwhelming majority of F-theory vacua are likely constructed from
Calabi-Yau manifolds that have one or more non-Higgsable clusters,
which may mean that symmetries are in fact generic and a large part of
the landscape.  More specifically, for Calabi-Yau threefolds the
complete set of toric bases $B_2$ and the larger set of ``semi-toric''
bases were analyzed in \cite{toric, WT-Hodge,
  Martini-WT}.  Out of all these bases, the vast majority had multiple
non-Higgsable clusters, largely consisting of $E_8, F_4,$ and $SU(2)
\times G_2$ clusters with minimal matter. For example, of more than
60,000 toric bases, only 16 (the toric del Pezzo and generalized del
Pezzo surfaces) lacked non-Higgsable clusters.  While the number of
distinct Calabi-Yau manifolds that can be constructed as (``tuned'')
elliptic fibrations (see {\it e.g.}  \cite{Johnson-Taylor}) over the
base is larger for the simpler bases that lack non-Higgsable clusters,
such bases also have smaller Hodge numbers and have moduli spaces that
are in some sense ``smaller''.  In general, bases with larger Hodge
numbers tend to have more non-Higgsable clusters, associated with
large non-Higgsable gauge groups with many factors of the factors
mentioned above.

For four-dimensional F-theory constructions, all indications are that
a similarly vast majority of all allowed threefold bases give
non-Higgsable clusters. Furthermore, those Calabi-Yau fourfolds with
many non-Higgsable clusters tend to be those that have large Hodge
numbers and therefore a larger number of expected stabilized flux
vacua.  It has been noted previously in the literature that
compactifications on fourfolds with large Hodge numbers give large
gauge groups \cite{Candelas-pr}.  Non-Higgsable clusters appear in other
approaches to string compactification as well as F-theory.  In
\cite{Anderson-WT}, the set of toric threefold bases $B_3$ was
constructed that give F-theory compactifications having dual heterotic
descriptions on smooth elliptically fibered Calabi-Yau threefolds with
smooth gauge bundles.  Of these constructions the majority (roughly
85\%) exhibited non-Higgsable clusters.  Furthermore, these
constructions correspond precisely to those that have the least
potential for non-Higgsable structure; the rest of the examples in a
much larger class of bases with a similar structure as $\P^1$ bundles
over a general toric base \cite{Halverson-WT} generically exhibit
non-Higgsable structures corresponding to singularities in the
Calabi-Yau threefold carrying enhanced gauge symmetries on the
heterotic side. 

Thus, from our current understanding it seems quite plausible that the
landscape is overwhelmingly dominated by vacua exhibiting symmetry.
In such a scenario, the kind of non-Higgsable QCD model which we
introduced in this paper may be a promising approach to realizing
observed physics in a natural way a generic points in the string landscape. More
broadly, non-Higgsable clusters may play an important role in
determining the structure of symmetries and light particles at low
energy scales.
Given the apparent genericity of non-Higgsable clusters in the
landscape, it is worth noting that they also may play other important
roles, such as for supersymmetry breaking
or dark matter phenomenology.

\sskip
\noindent \emph{Non-Higgsable QCD and Model Building}
\vspace{.1cm}

The existence of an unbroken QCD sector is an experimental fact that
also motivates the kind of scenario we have considered in this paper.
This is a particularly important fact since if QCD were broken at a
scale much higher than $\lqcd\simeq 200\,$ MeV, quark confinement
would not occur and protons would cease to exist, changing physics
drastically in a way not amenable to life as we know it. In the
standard model the existence of an unbroken QCD sector is a simple
artifact of the theory; it has no colored scalar. On the other hand,
in four-dimensional theories with $\cN=1$ supersymmetry the existence
of an unbroken QCD sector must be accounted for, as such theories
necessarily have colored scalars. This constrains model-building.

For example, in the minimal supersymmetric standard model (MSSM) there
are supersymmetric color-breaking (nearly) flat directions; see
\cite{GKM} for a classification and \cite{Veronese} for an analytic
study of the electroweak vacuum moduli space. The possible existence
of color-breaking vacua constrains the soft breaking parameters. For
example, the so-called A-terms are scalar trilinear soft breaking
terms, and there are upper bounds on some of these parameters
necessary for the absence of dangerous color breaking vacua; see for
example \cite{CCBConstraints-I,CCBConstraints-II} and references
therein. The term for the top quark $A_t$ appears in loop corrections
to the mass of the Higgs boson, and $A_t$ must often be large in order
to account for its observed value. Together, these two constraints can
place significant tension on models; see e.g. \cite{CohenWacker} for
one recent account in the CMSSM.

In the scenario we have proposed, the existence of an unbroken QCD
sector is a natural possibility when model-building with non-Higgsable
clusters in F-theory. As we have emphasized, $SU(3)$ and $SU(2)$ are
the only $SU(n)$ groups that may be realized through a non-Higgsable
cluster. We have considered the possibility that $SU(3)_{QCD}$ is
realized as a non-Higgsable cluster, which requires a non-perturbative
seven-brane realized by a type $IV$ fiber. We looked in particular
detail at one geometry, obtained from the intersection of a type $IV$
and a type $III$ fiber.

There are a number of interesting model-building implications of
non-Higgsable clusters.  First, conventional $SU(5)$ grand unification
cannot be realized at a generic point in the moduli space of a
compactification with a non-Higgsable cluster; specialization in
moduli space is necessary. $E_6$ or $SO(10)$ grand unification may be
possible with a non-Higgsable cluster, however, as discussed in
section \ref{sec:SM}, with $SO(10)$ also requiring specialization.

Second, though we have classified in broad strokes the possible
realizations of a non-Higgsable QCD sector in F-theory, we have not
performed a detailed analysis of all of the geometries. Such an
analysis may imply the existence of potentially interesting exotics
that are experimentally allowed and potentially discoverable at the
LHC. Exotics are known to be a feature in many corners of the
landscape and their existence can sometimes be related to string
consistency conditions that do not have a simple analog in quantum
field theory; see {\it e.g.} \cite{AnomalyNucleation,
  ExoticImplications} for a study of how stringy constraints can imply
the existence of quasichiral electroweak exotics.  More generally,
while anomalies can act as a low-energy window on constraints that
must be satisfied by any UV completion of a quantum theory, the
geometry of F-theory vacua may place further specific constraints on the set
of low-energy fields and interactions available in the 4D theory in
ways that are not yet understood from the low-energy point of view.
Such considerations have led to a productive line of inquiry in the context of
six-dimensional theories  (see {\it e.g.} \cite{KMT-II, Seiberg-WT});
an initial investigation of possible constraints that may arise in
this fashion for four-dimensional theories was carried out in
\cite{Grimm-Taylor}.

Finally, while our scenario requires the existence of a non-Higgsable
QCD sector, we leave open the possibility that there exist other
non-Higgsable gauge factors. As mentioned above, an interesting
possibility is that of a non-Higgsable hidden sector, which may give
rise to interesting dark matter or supersymmetry breaking
phenomenology. Also, given that $SU(2)$ and $U(1)$ are gauge factors
that may also be non-Higgsable, it would be interesting to study the
possibility of a non-Higgsable electroweak sector. At first glance
this seems to be ruled out experimentally, but given the successes of
radiative electroweak symmetry breaking in other theories it may be
worth considering. If possible it would likely, as in the MSSM, depend
heavily on the details of supersymmetry breaking and renormalization
group flow.  In one sense it would also appeal to minimality; only ten
individual group factors may be realized by non-Higgsable
seven-branes, and it so happens that the three factors of lowest
dimension are precisely $SU(3)$, $SU(2)$, and $U(1)$.

\vspace{.5cm}
\noindent {\bf Acknowledgments.} \\
\noindent We thank Lara Anderson, Mirjam Cveti{\v c}, Keshav Dasgupta,
Tony Gherghetta, James Gray, Tom Hartman, Kristan Jensen, Sam Johnson,
Denis Klevers, Paul Langacker, Liam McAllister, Paul McGuirk, Dave
Morrison, Daniel Park, Joe Polchinski, and Jesse Thaler for useful
conversations. J.H. thanks J.L. Halverson for her kind support and
constant encouragement. The research of J.H. was supported in part by
the National Science Foundation under Grant No. PHY11-25915.  The
research of W.T.\ is supported by the U.S.\ Department of Energy under
grant Contract Number DE-SC00012567. J.L.S. is supported by DARPA,
fund no. 553700, and is the Class of 1939 Professor in the Schools of
Arts and Sciences of the University of Pennsylvania; he gratefully
acknowledges the generosity of the Class of 1939. A.G. is supported by
NSF Research Traning Group Grant GMS-0636606. We thank the Aspen
Center for Physics for hospitality and partial support by the National
Science Foundation Grant No. PHYS-1066293.  We also thank the Simons
Center for Geometry and Physics for hospitality while this work was
carried out.

\begin{table}
  \centering
  \begin{tabular}{ccc}
    Junction & Dynkin Labels &  \shs Representation \shs \\ \hline \hline
    $ (-1, -1, 1, 1, 0, -1, 1, 0) $ & $ (0, 0, 0) $ & Singlet\\
    \hline
    $ (-1, 0, 1, 0, -1, -1, 0, -1) $ & $ (0, 0, 0) $ & Singlet\\
    \hline
    $ (-1, 0, 1, 1, 0, 0, 0, -1) $ & $ (1, 0, 1) $ & $(\textbf{3},\textbf{2})$ \\
    $ (-1, 0, 1, 1, 0, -1, 0, 0) $ & $ (-1, 1, 1) $ & \\
    $ (-1, -1, 0, 0, 0, 0, 0, -1) $ & $ (1, 0, -1) $ & \\
    $ (-1, 0, 1, 1, -1, -1, 1, 0) $ & $ (0, -1, 1) $ & \\
    $ (-1, -1, 0, 0, 0, -1, 0, 0) $ & $ (-1, 1, -1) $ & \\
    $ (-1, -1, 0, 0, -1, -1, 1, 0) $ & $ (0, -1, -1) $ & \\
    \hline
    $ (-1, -1, 0, 1, 1, 0, 0, 0) $ & $ (0, 1, 0) $ & $(\overline{\textbf{3}},\textbf{1})$\\
    $ (-1, -1, 0, 1, 0, 0, 1, 0) $ & $ (1, -1, 0) $ & \\
    $ (-1, -1, 0, 1, 0, -1, 1, 1) $ & $ (-1, 0, 0) $ & \\
    \hline
    $ (-1, 0, 0, 0, 0, 0, -1, -1) $ & $ (0, 1, 0) $ & $(\overline{\textbf{3}},\textbf{1})$\\
    $ (-1, 0, 0, 0, -1, 0, 0, -1) $ & $ (1, -1, 0) $ & \\
    $ (-1, 0, 0, 0, -1, -1, 0, 0) $ & $ (-1, 0, 0) $ & \\
    \hline
    $ (-1, 0, 0, 1, 0, 0, 0, 0) $ & $ (0, 0, 1) $ & $(\textbf{1},\textbf{2})$\\
    $ (-1, -1, -1, 0, 0, 0, 0, 0) $ & $ (0, 0, -1) $ & \\
    \hline 
    \hline
    $ (0, 0, 1, 1, 1, 0, 0, 0) $ & $ (0, 1, 1) $ & $(\overline{\textbf{3}},\textbf{2})$\\
    $ (0, 0, 1, 1, 0, 0, 1, 0) $ & $ (1, -1, 1) $ & \\
    $ (0, -1, 0, 0, 1, 0, 0, 0) $ & $ (0, 1, -1) $ & \\
    $ (0, 0, 1, 1, 0, -1, 1, 1) $ & $ (-1, 0, 1) $ & \\
    $ (0, -1, 0, 0, 0, 0, 1, 0) $ & $ (1, -1, -1) $ & \\
    $ (0, -1, 0, 0, 0, -1, 1, 1) $ & $ (-1, 0, -1) $ & \\
    \hline
    $ (0, 1, 1, 0, 0, 0, -1, -1) $ & $ (0, 1, 1) $ & $(\overline{\textbf{3}},\textbf{2})$\\
    $ (0, 1, 1, 0, -1, 0, 0, -1) $ & $ (1, -1, 1) $ & \\
    $ (0, 0, 0, -1, 0, 0, -1, -1) $ & $ (0, 1, -1) $ & \\
    $ (0, 1, 1, 0, -1, -1, 0, 0) $ & $ (-1, 0, 1) $ & \\
    $ (0, 0, 0, -1, -1, 0, 0, -1) $ & $ (1, -1, -1) $ & \\
    $ (0, 0, 0, -1, -1, -1, 0, 0) $ & $ (-1, 0, -1) $ & \\
    \hline
    $ (0, 0, 1, 0, 0, 0, 0, -1) $ & $ (1, 0, 0) $ & $(\textbf{3},\textbf{1})$\\
    $ (0, 0, 1, 0, 0, -1, 0, 0) $ & $ (-1, 1, 0) $ & \\
    $ (0, 0, 1, 0, -1, -1, 1, 0) $ & $ (0, -1, 0) $ & \\
    \hline
    $ (0, -1, 0, 1, 1, 0, 1, 1) $ & $ (0, 0, 0) $ & Singlet \\ 
    \hline \hline
    $ (0, 0, 0, 0, 1, 1, -1, -1) $ & $ (1, 1, 0) $ & $SU(3)$ Pos. Roots\\
    $ (0, 0, 0, 0, 1, 0, -1, 0) $ & $ (-1, 2, 0) $ & \\
    $ (0, 0, 0, 0, 0, 1, 0, -1) $ & $ (2, -1, 0) $ & \\
    \hline
    $ (0, 1, 1, 1, 0, 0, 0, 0) $ & $ (0, 0, 2) $ & $SU(2)$ Pos. Roots\\ \hline \hline
  \end{tabular}
  \caption{Displayed are half the string junctions in an elliptic
    surface near the $IV$-$III$ collision; group theoretically, these
    are the $36$ positive roots in an adjoint of $E_6$. We also
    display their representation under $SU(3)\times SU(2)$ and
    corresponding Dynkin labels; in the latter, the first two entries
    and last entry are the $SU(3)$ and $SU(2)$ Dynkin labels,
    respectively. The first sixteen junctions end on the extra brane
    and are in the same $SU(3)\times SU(2)$ representations as a
    generation of standard model fermions. 
The relationship of this data
    to the matter spectrum at the $IV$-$III$ collision is described in
    detail in the text; they are
    related but not identical, since we have only performed a partial
    deformation analysis.}
\label{table:IV-III Reps}  
\end{table}

\end{document}